\documentclass[12pt]{article}
\pdfoutput=1
\usepackage{epsfig}
\usepackage{comment}
\usepackage{latexsym}
\usepackage{color}
\usepackage{amsmath}
\usepackage{hyperref}

\newcommand{\mysquare}[0]{\raise-.2ex\hbox{{\Large$\Box$}}}
\def\lsim{\mathrel{\rlap {\raise.5ex\hbox{$ < $}}
{\lower.5ex\hbox{$\sim$}}}}
\def\gsim{\mathrel{\rlap {\raise.5ex\hbox{$ > $}}
{\lower.5ex\hbox{$\sim$}}}} \topmargin -1.5cm \textheight=22.5cm \textwidth=16.5cm
\catcode`\@=11
\newcount\hour
\newcount\minute
\newtoks\amorpm
\hour=\time\divide\hour by60 \minute=\time{\multiply\hour by60 \global\advance\minute by-\hour}
\edef\standardtime{{\ifnum\hour<12 \global\amorpm={am}%
        \else\global\amorpm={pm}\advance\hour by-12 \fi
        \ifnum\hour=0 \hour=12 \fi
        \number\hour:\ifnum\minute<10 0\fi\number\minute\the\amorpm}}
\edef\militarytime{\number\hour:\ifnum\minute<10 0\fi\number\minute}
\def\draftlabel#1{{\@bsphack\if@filesw {\let\thepage\relax
   \xdef\@gtempa{\write\@auxout{\string
      \newlabel{#1}{{\@currentlabel}{\thepage}}}}}\@gtempa
   \if@nobreak \ifvmode\nobreak\fi\fi\fi\@esphack}
        \gdef\@eqnlabel{#1}}
\def\@eqnlabel{}
\def\@vacuum{}
\def\draftmarginnote#1{\marginpar{\raggedright\scriptsize\tt#1}}
\def\draft{\oddsidemargin -.2truein
        \def\@oddfoot{\sl preliminary draft \hfil
        \rm\thepage\hfil\sl\today\quad\militarytime}
        \let\@evenfoot\@oddfoot \overfullrule 3pt
        \let\label=\draftlabel
        \let\marginnote=\draftmarginnote
   \def\@eqnnum{(\theequation)\rlap{\k

 ern\marginparsep\tt\@eqnlabel}%
\global\let\@eqnlabel\@vacuum}  }

\newcommand{\be}[0]{\begin{equation}}
\newcommand{\ee}[0]{\end{equation}}
\newcommand{\ba}[0]{\begin{eqnarray}}
\newcommand{\ea}[0]{\end{eqnarray}}

\def\bs{\begin{subequations}}
\def\es{\end{subequations}}

\def\thebibliography#1{%
\vskip 0.5cm \centerline{\bf \Large References}
\list{%
[\arabic{enumi}]}{\settowidth\labelwidth{[#1]} \leftmargin\labelwidth \advance\leftmargin\labelsep
\usecounter{enumi}}
\def\newblock{\hskip .11em plus .33em minus .07em}
\sloppy\clubpenalty4000\widowpenalty4000 \sfcode`\.=1000\relax}

\renewcommand{\theequation}{\arabic{section}.\arabic{equation}}

\renewcommand{\section}{\setcounter{equation}{0}\@startsection
{section}{1}{0mm}{-\baselineskip}{0.5\baselineskip} {\normalfont\Large\bfseries}}

\renewcommand{\subsection}{\@startsection
{subsection}{2}{0mm}{-\baselineskip}{0.5\baselineskip} {\normalfont\large\bfseries}}

\renewcommand{\subsubsection}{\@startsection
{subsubsection}{3}{0mm}{-\baselineskip}{0.5\baselineskip} {\normalfont\normalsize\slshape}}

\usepackage{amssymb,amsfonts}
\usepackage{graphicx}

\newcommand{\bea}{\begin{eqnarray}}
\newcommand{\eea}{\end{eqnarray}}
\newcommand{\dis}{\displaystyle}

\newcommand{\Z}{\mathbb{Z}}

\renewcommand{\O}{{\cal O}}

\newcommand{\sign}{{\rm sign}}
\newcommand{\abs}{|}

\newcommand{\ie}{{\em i.e. }}
\newcommand{\where}{\mbox{where}}
\newcommand{\with}{\mbox{with}}

\renewcommand{\and}{\mbox{and}}

\newcommand{\F}{{\cal F}}
\newcommand{\N}{{\cal N}}
\newcommand{\M}{{\cal M}}
\newcommand{\A}{{\cal A}}

\newcommand{\K}{{\cal K}}

\renewcommand{\F}{{\cal F}}

\renewcommand{\o}{\overset{\circ}}
\newcommand{\oo}{\overset{\circ\circ}}

\newcommand{\tg}{\tilde g}
\newcommand{\tk}{\tilde k}
\newcommand{\tm}{\tilde m}
\newcommand{\tn}{\tilde n}

\renewcommand{\b}{\bar}

\begin{document}
\begin{titlepage}
\begin{flushright}
CPHT--RR009.0210,
LPTENS--09/32,
February 2010
\end{flushright}

\vspace{2mm}

\begin{centering}
{\bf \huge Superstring cosmology for $\N_4 $=$ 1$$ \rightarrow$$ 0$ superstring vacua}\\

\vspace{8mm}
 {\Large John Estes$^{1,2}$, Costas Kounnas$^{2}$ and Herv\'e Partouche$^1$}

\vspace{2mm}

$^1$ Centre de Physique Th\'eorique, Ecole Polytechnique,$^\dag$
\\
F--91128 Palaiseau cedex, France\\
{\em John.Estes@cpht.polytechnique.fr}\\
{\em Herve.Partouche@cpht.polytechnique.fr}

\vspace{2mm}

$^2$ Laboratoire de Physique Th\'eorique,
Ecole Normale Sup\'erieure,$^\ddag$ \\
24 rue Lhomond, F--75231 Paris cedex 05, France\\
{\em  Costas.Kounnas@lpt.ens.fr}

\vspace{5mm}

{\bf\Large Abstract}

\end{centering}
\vspace{4mm}

\noindent
We study the cosmology of perturbative heterotic superstring theory during the radiation-like era for semi-realistic backgrounds with initial $\N=1$ supersymmetry.  This analysis is valid for times after the Hagedorn era (or alternatively inflation era) but before the electroweak symmetry breaking transition.  We find an attraction to a radiation-like era with the ratio of the supersymmetry breaking scale to temperature stabilized.  This provides a dynamical mechanism for setting the supersymmetry breaking scale and its corresponding hierarchy with the Planck scale.  For the internal space, we find that orbifold directions never decompactify, while toroidal directions may decompactify only when they are wrapped by certain geometrical fluxes which break supersymmetry.  This suggests a mechanism for generating spatial directions during the radiation-like era.  Moreover, we show that certain moduli may be stabilized during the radiation-like era with masses near the supersymmetry breaking scale. In addition, the moduli do not dominate at late times, thus avoiding the cosmological moduli problem.


\vspace{3pt} \vfill \hrule width 6.7cm \vskip.1mm{\small \small \small
\noindent
 $^\dag$\ Unit{\'e} mixte du CNRS et de l'Ecole Polytechnique,
UMR 7644.}\\
$^\ddag$\ Unit{\'e} mixte  du CNRS et de l'Ecole Normale Sup{\'e}rieure associ\'ee \`a
l'Universit\'e Pierre et Marie Curie (Paris 6), UMR 8549.

\end{titlepage}
\newpage
\setcounter{footnote}{0}
\renewcommand{\thefootnote}{\arabic{footnote}}
 \setlength{\baselineskip}{.7cm} \setlength{\parskip}{.2cm}

\setcounter{section}{0}


\section{Introduction}
\label{intro}

One approach to modern cosmology at the fundamental level focuses on the study of time-dependent backgrounds that implement a period of inflation. Such a period of accelerated expansion provides explanations of the homogeneity, isotropy, flatness, large size and entropy of the Universe \cite{inflation}. It also gives an origin of the nearly scale-invariant spectrum of primordial cosmological fluctuations. Originally introduced in the context of field theory, a natural goal is to implement this scenario in string theory by taking into account both perturbative and non-perturbative quantum corrections and looking for solutions which are de-Sitter like \cite{stringinflation}.  Often in such inflationary scenarios, one encounters the cosmological moduli problem \cite{cosmomodprob}, where at intermediate times the Universe is not thermal but rather dominated by the energy stored in massive moduli.  Their eventual decay can lead to problems such as excess entropy production.

A drawback of the inflationary paradigm is that it does not explain how to resolve the initial Big Bang singularity.  Another approach, stringy in origin, is to use the Hagedorn transition\cite{AtickWitten,AKADK,Hagedorn}  that occurs at the ultra high temperature $T_H$  as an alternative approach to inflation \cite{hagedorncosmology,Kaloper:2007pw}.  However, it is again often difficult to maintain analytical control when trying to implement dynamically such a phase transition. In order to address this problem, string backgrounds have been recently constructed which do not present a breakdown of the canonical ensemble description, due to a duality involving the Euclidean time circle \cite{GravFluxes,MassSusy}. A study of the cosmological behavior for a special example can be found in \cite{hybrid}.

In both approaches, inflation or the alternative stringy mechanism at the Hagedorn transition,
there is a problem associated to connection of these early times to the standard matter dominated cosmology.  For example, one needs to stabilize moduli after which one often runs into the cosmological moduli problem \cite{cosmomodprob}.  In the present work, we find a radiation-like evolution which connects the Hagedorn era (ending at a time $t_E$) to the electro-weak era (starting at a time $t_W$), where the standard model particles gain their masses.  When considering this intermediate epoch ($t_E < t < t_W$), it is possible to parameterize our ignorance of the earlier evolution by considering arbitrary initial boundary conditions (IBC) at $t_E$. From the outset at $t_E$, in a perturbative approach, one considers a string background defined by a two-dimensional conformal field theory and computes the quantum and thermodynamical properties of a space filling  thermalized gas of string states \cite{stringgascosmo}. All statistical properties at thermal equilibrium are derived from the underlying microscopic theory: First and second laws of thermodynamics together with state equations.  The gas back-reacts on the classically static background and in certain cases, which we characterize, leads to a well defined cosmological evolution.

Following \cite{KP2,cosmo1,cosmo2,Bourliot:2009cx}, we implement these ideas within the framework of two-dimensional conformal field theories that define tree level superstring compactifications in four-dimensional flat space, with either $\N = 2$ or $\N = 1$ supersymmetry spontaneously broken to $\N=0$.
This breaking is introduced by geometrical fluxes in the internal space \cite{SS,Rohm,KouPor,RostKounnas,GeoFluxes,OpenFluxes}. Finite temperature $T$ is switched on by considering the Euclidean version of the model where the time is compactified on a circle of perimeter $\beta=2\pi R_0$.
Again, an appropriate geometrical flux along the compact Euclidean time is introduced to implement (anti-)periodic boundary conditions for (fermions) bosons. The advantage of using geometrical fluxes, is that our analysis is exact in the string scale $\alpha^\prime$.
At the one-loop level, the partition function (or free energy) is non-vanishing and implies a pressure and an energy density that induce the above mentioned cosmological evolution.
A special role is played by the supersymmetry breaking scales; namely the temperature $T$ and the universal no-scale-modulus $\, \Phi$, which appears in all $\N = 1$ effective supergravity theories and defines the supersymmetry breaking scale $M(\Phi)$ \cite{Noscale, StringyNoscale} at  zero temperature.

The moduli that are participating in the spontaneous breaking of supersymmetry are running away, namely the supersymmetry breaking scale is evolving in time, while their ratios are stabilized \cite{cosmo1,cosmo2}. The analysis of the dynamics of the spectator moduli $\mu_I$ which are not involved in the supersymmetry breaking was initiated in \cite{BEKP,Partouche:2010cq}. There, it was found that for $\mu_I > T$ and $M$, the one-loop partition function (to all orders in the string length) or equivalently the free energy density ${\cal F}$ takes the following form during the intermediate era, \ie between the time $t_E$ (the Hagedorn exit time) and the time $t_W$ (the electroweak phase transition),
\be
\label{moduli}
{\cal F}(T,M; \mu_I)={\cal F}(T,M)+{\cal O}\left[~{\rm exp}\left(- {\mu_I\over T}\right),~{\rm exp}\left(- {\mu_I\over M}\right)~\right]\,.
\ee
As a result, it turns out that the $\mu_I$'s are either dynamically stabilized at their self-dual (or enhanced symmetry) points, with masses of order the supersymmetry breaking scale, or frozen due to the expansion of the universe.  The fact that the back-reaction of a gas of strings can stabilize a radius at its self dual point was initially discussed in \cite{Watson:2004aq,Patil:2004zp} (see also \cite{stringgascosmo,Greene:2007sa}).  Here we see that once supersymmetry is broken, the stabilization also arises from purely thermal/quantum effects.  In particular, once the supersymmetry breaking scale is stabilized, as discussed below, the quantum effects will be sufficient to stabilize the radius with a mass around the supersymmetry breaking scale.

For a large class of models, the time-trajectories of $M(t)$ and $T(t)$ are such that the ratio $M/T$ is stabilized to some model dependent constant.  The external space-time evolution is \emph{attracted} to a Friedmann-Lema\^{\i}tre-Robertson-Walker (FLRW) type cosmology corresponding to a ``Radiation-like Dominated Solution" in four dimensions (RDS$^4$) \cite{Bourliot:2009na}.  It satisfies
 \be
 \label{RDS}
{\rm RDS}^4~:~~~~~ M(t) \propto T(t) \propto {1\over a(t)} \propto e^{4 \phi(t)} \propto {1\over \sqrt{t}} ~,~~~~{\rm for} ~~~~t_E\le t\le t_W\; ,
 \ee
where $a$ is the spatial scale factor and $\phi$ is the four-dimensional dilaton.\footnote{As a necessary and sufficient consistency requirement, we note that in this intermediate cosmological regime, the smallness of the space-time curvature scales ${H^2=(\dot a}/a)^2$ and $\dot H$, the dilaton scales $\dot\phi^2$, $\ddot \phi$, and the evolving internal radii scales, $({\dot R}_I/R_I)^2$, $\ddot R_I/R_I$, is guaranteed thanks to the ``attractor mechanism" towards the RDS$^4$.  In particular, they are all decreasing at late cosmological times together with $e^\phi$ so that our quasi-static and perturbative approximations become better and better as time passes.}
Strictly speaking, the attractor (\ref{RDS}) {\em does not} describe a radiation dominated universe but mimics it. This is due to the fact that the time-dependent state equation of the thermal gas of string modes does not converge to Stefan's law. It is only by taking into account the contribution of the coherent motion  of the modulus $M(t)$ that the {\em total} energy density and pressure in the Universe satisfy  $\rho_{\rm tot} = 3 P_{\rm tot}$ during the intermediate era.

Along the radiation-like evolution, the supersymmetry breaking scale is \emph{not stabilized} but rather falls in time proportional to the temperature.  It is exactly this fact which allows us to evade the cosmological moduli problem.  The masses of the moduli are related to the supersymmetry breaking scale and are thus also falling in time.  The net effect is that the energy stored in the moduli dilutes faster than the thermal energy, and the evolution always remains radiation-like.  In addition, the supersymmetry breaking scale $M(t)$ is naturally following the temperature scale $T(t)$, thus the thermal relic density of supersymmetric particles is naturally low and we avoid the cosmological gravitino problem.  At the end of the intermediate era, $t\lsim t_W$, it is still clearly necessary to stabilize the supersymmetry breaking scale and also the evolution of the moduli masses. Fortunately at least one known mechanism already exists, namely radiative electro-weak symmetry breaking which has been shown in certain cases to stabilize the supersymmetry breaking scale around the electro-weak scale \cite{NoscaleTSR}. This provides a natural explanation of the hierarchy between the string scale and the supersymmetry breaking scale and is further discussed in the conclusion.

In Sect. \ref{FZa}, we first calculate the free energy density at one-loop for the simplest orbifold models with initial $\N=2$ and $\N=1$ supersymmetry, in order to uncover general behavior. We then discuss how these features apply to more realistic models, such as those of \cite{Assel:2010wj}.  In Sect. \ref{radera}, we study the resulting induced cosmologies. In Sect. \ref{T>M},  we find that toroidal-like internal directions with geometrical fluxes can be dynamically decompactified. This shows the space-time dimension can change during the intermediate cosmological era.  As a starting point, we perform these analysis while keeping fixed the remaining spectator moduli $\mu_I$ at values much larger than $T(t_E)$ and $M(t_E)$.  In Sect. \ref{Rdyna}, we relax this restriction and analyze the dynamics of the spectator moduli in the presence of the induced thermal effective potential.  In Sect. \ref{Summ}, we give a summary of our results, while in Sect. \ref{Conc} we give our conclusions.  In appendix A, we study the simpler case where supersymmetry is spontaneously broken by thermal effects only.  There, we derive the very well known Stefan's law and black body physics in arbitrary dimension $D$ from the effective supergravity theory of string theory models at finite temperature.


\section{Free energy density}
\label{FZa}

In this section, our aim is to derive the canonical ensemble 1-loop free energy associated to Euclidean string backgrounds. We first consider the simplest cases in order to understand the generic behavior of the free energy density.  We then discuss how our results generalize to other backgrounds, including the semi-realistic models of \cite{Assel:2010wj}.  For the computation to be sensible, we regularize the spatial volume by putting it in a ``large box''. By ``box'', it is understood a squared torus $T^3$ with radius $R_{\rm box}$, while ``large'' means the sum on the Kaluza-Klein (KK) states is replaced by a continuous integral. We analyze three kinds of Euclidean models: In case (I), we take
\be
\label{backg1}
{\rm (I)\; :} \qquad S^1(R_0)\times T^{3}(R_{\rm box})\times S^1(R_4)\times S^1\times {T^4\over \Z_2}\, ,
\ee
with flux in the directions 0 and 4. This corresponds to a background with $\N = 2 \rightarrow 0$ at finite temperature. Next, is case (II), the internal flux wraps an orbifold direction,
\be
\label{backg2}
{\rm (II)\; :} \qquad S^1(R_0)\times T^{3}(R_{\rm box})\times {S^1(R_4)\times T^3\over \Z_2}\times T^2\, .
\ee
Case (III) is a model with $\N = 1 \rightarrow 0$,
\be
\label{backg3}
{\rm (III)\; :} \qquad S^1(R_0)\times T^{3}(R_{\rm box})\times {S^1(R_4)\times T^5\over \Z_2\times \Z_2}\, , \hskip0.4in
\ee
with again flux in the directions 0 and 4.

 In these models, the FLRW scale factor $a$ scales as $R_{\rm box}$. The supersymmetry is spontaneously broken at zero temperature, with mass shift scale $M\propto 1/R_4$, by the presence of internal flux. The computation is specialized to the intermediate cosmological era where $T\propto 1/R_0$ and $M$ are much smaller than the Hagedorn temperature \ie
\be
R_0\quad \and\quad  R_4\gg 1\, ,
\ee
(and above the electroweak scale in realistic models).
We also suppose that the remaining moduli associated to the internal directions $X^I$ ($I=5,\dots,9$) introduce scales that are very large compared to $T$ and $M$. Quantitatively, this means the internal radii satisfy
\be
\label{hyp}
{1\over R_0}\ll R_I\ll R_0 \qquad \and \qquad {1\over R_4}\ll R_I\ll R_4\, , \qquad I=5,\dots, 9\, .
\ee
At this point, the reader interested in the induced cosmologies can switch to Sects \ref{radera} and \ref{T>M}, where the attractor solutions are found. It will be shown in Sect. \ref{Rdyna} that the hypothesis (\ref{hyp}) is consistent and always reached dynamically.
Before we enter into our subject, let us comment that our formalism and ideas are presented in appendix A in much simpler models where supersymmetry is spontaneously broken by thermal effects only. There, the computation of the free energy density, the induced cosmology and the moduli stabilization for this simple case are presented.


\subsection{$\N$ = 2 $\to$ 0 models: Cases (I) and (II)}
\label{FZ2}

In Einstein frame, the free energy density $\F$ is related to the Euclidean partition function $Z$ as
\be
\label{FZ}
\F=-{Z\over (e^{-\phi}\beta)\, (e^{-3\phi}V_{\rm box})}\, ,
\ee
where $\beta=2\pi R_0$, $V_{\rm box}=(2\pi R_{\rm box})^3$ and $\phi$ is the dilaton in four dimensions. In case (I), Eq. (\ref{backg1}), its derivation for the heterotic string can be found in \cite{cosmo1, BEKP} and is summarized in appendix B. Let us give more details in case (II), Eq. (\ref{backg2}).

To compute the Euclidean partition function, the temperature is introduced by imposing non-trivial boundary conditions.  This is achieved by inserting a phase in the $\Gamma_{(1,1)}$ lattice of zero modes of $S^1(R_0)$. For the pure KK states, this phase is $(-1)^a$, where $a$ is the fermionic number.  Requiring modular invariance then fixes the form for general states.  Similarly, for spontaneous supersymmetry breaking by an internal flux along $S^1(R_4)$, one inserts in the associated $\Gamma_{(1,1)}$ lattice a phase determined by a choice of R-symmetry charge $a+\b Q_4$.
The resulting partition function is
\ba
\nonumber Z=\dis R_0R_{\rm box}^3&&\dis \!\!\!\!\!\!\!\! \! \!\int_F{d\tau_1d\tau_2\over 2\tau_2^3}\dis{1\over 2}\sum_{H,G}{1\over 2}\sum_{a,b}(-)^{a+b+ab}{\theta[^a_b]^2\theta[^{a+H}_{b+G}]\theta[^{a-H}_{b-G}]\over \eta^4}\sum_{n_0,\tm_0}e^{-{\pi R_0^2\over \tau_2}\abs \tm_0+n_0\tau\abs^2}(-)^{a\tm_0+bn_0+\tm_0n_0}\\
&&\dis \!\!\!\!\!\!\!\! \! \!{1\over 2}\sum_{\b\gamma,\b\delta}\b\theta[^{\b\gamma}_{\b\delta}]^8\, {1\over 2}\sum_{\b\gamma',\b\delta'}\b\theta[^{\b\gamma'}_{\b\delta'}]^6\b\theta[^{\b\gamma'+H}_{\b\delta'+G}]\b\theta[^{\b\gamma'-H}_{\b\delta'-G}]\, Z_{(1,1)}^{(a+\b Q_4,b+\b L_4)}[^H_G]\, Z_{(3,3)}^{(0,0)}[^H_G]\, {\Gamma_{(2,2)}\over \eta^4\bar\eta^{20}}\, ,
\ea
where $Z^{(0,0)}_{(n,n)}$ is the standard contribution associated to the $\Z_2$-twist on $T^n$ given in (\ref{Tn/Z2}) and the orbifold block $Z_{(1,1)}^{(a+\b Q_4,b+\b L_4)}[^H_G]$ computed in appendix C is
\be
\label{Z11}
Z_{(1,1)}^{(a+\b Q_4,b+\b L_4)}[^H_G]=\left\{
\begin{array}{ll}
\dis {R_4\over  \sqrt{\tau_2}\eta\b\eta}\sum_{n_4\tm_4}e^{-{\pi R_4^2\over \tau_2}\abs \tm_4+n_4\tau\abs^2}(-)^{(a+\b Q_4)\tm_4+(b+\b L_4)n_4+\b\epsilon_4 \tm_4n_4}&\dis \mbox{for } [^H_G]\equiv [^0_0],\\
\dis 2\sqrt{{\eta\b\eta\over \theta[^{1-H}_{1-G}]\b\theta[^{1-H}_{1-G}]}}\; \delta_{(a+\b Q_4)G+(b+\b L_4)H+\b\epsilon_4 HG,0\, mod\, 2}&\dis\mbox{for } [^H_G]\not\equiv[^0_0].
\end{array}
\right.
\ee
In this expression, a generic choice of charge $\b Q_4$ is a linear sum of gauge group right moving lattice charges and/or orbifold twists, which can be represented as
\be
\label{charges}
\b Q_4=\b\eta \b\gamma+\b\eta'\b\gamma'+\b\eta^{\prime\prime}H\; ,\qquad \b L_4=\b\eta \b\delta+\b\eta'\b\delta'+\b\eta^{\prime\prime}G\; ,\qquad \b\epsilon_4=1-\b\eta-\b\eta'\, ,
\ee
where $\b\eta, \b\eta', \b\eta^{\prime\prime}$ are constants equal to 0 or 1. $\b L_4$ and $\b \epsilon_4$ are determined by modular invariance by considering the transformations $\tau\to -1/\tau$ and $\tau\to \tau+1$, respectively.

Clearly, the contribution of the sector $[^H_G]\equiv[^0_0]$ is identical for the backgrounds (I) and (II) and its computation in the intermediate cosmological era where $T,M\ll T_H$ is summarized in appendix B. The result involves the untwisted ``light" states only, as compared to $T_H$. Quantitatively, the substantial contributions arise from the massless modes together with their KK towers associated to the large radii $R_0$ and $R_4$, as compared to the string length. Exponentially suppressed terms in $R_0$ and/or $R_4$ can be neglected.

The situation is different for the contributions of the sectors $[^H_G]\not\equiv[^0_0]$. For the background (I), the dependence in $R_4$ is similar to the one arising in the sector $[^0_0]$ (see appendix B).  This is due to the fact that the orbifold acts on directions orthogonal to $S^1(R_4)$. However, this is not true anymore in case (II), where these sectors are independent of $R_4$, as can be seen in the second line of Eq. (\ref{Z11}).  The result is that these contributions are similar to the ones that would arise in a pure thermal case \ie with no spontaneous breaking of supersymmetry arising from the internal space,
and discussed in appendix A. The only difference between the $[^H_G]\not\equiv[^0_0]$ sector in  (\ref{Z11}) and its analog in the pure thermal case is the Kronecker symbol $\delta$ that may eliminate some sub-sectors. More specifically, for a choice of charge $\b Q_4$ in (\ref{charges}) with $\b\eta=0$, the $\delta$-symbol cuts contributions which are already vanishing (they all contain a $\theta[^1_1]$ factor): We are back to a standard derivation of the pressure arising in the pure thermal case. On the contrary, for $\b\eta=1$, the $\delta$-symbol eliminates the sectors $[^{\b\gamma}_{\b\delta}]=[^{1-H}_{1-G}]$. This has no consequence in the untwisted sector $[^H_G]\equiv [^0_1]$ which is  exponentially small in the regime $R_0\gg 1$. However, in the twisted sectors $[^H_G]\equiv [^{\,1}_G]$, the subtraction of ${1\over 2}\b\theta[^{\b\gamma}_{\b\delta}]^8={1\over 2}\b\theta[^{1-H}_{1-G}]^8$ from the $E_8$ lattice contribution is equivalent to replacing its $\b q$-expansion in the pure thermal case as $1+ \O(\b q)\to {1\over 2}+ \O(\b q)$.

Altogether, the 1-loop free energy density $\dis \F$ is given by
\be
\label{F2}
{\rm (II)\; :} \qquad \F=-T^4\, \Big( n^u_T\, f_T(z)+n^t_T\left(1-{\b\eta\over 2}\right) c_4+n^u_V\, f_V(z)\Big) \, ,
\ee
where
\be
e^z:={R_0\over R_4}={M\over T}\; ,\qquad T={1\over 2\pi R_0 \, e^{-\phi}}\; ,\qquad M={1\over 2\pi R_4 \, e^{-\phi}}
\ee
and $c_4=\dis {\pi^2\over 48}$ is defined in Eq. (\ref{Zstef}).
$n^u_T$ ($n^t_T$) is the number of untwisted (twisted) massless boson-fermion pairs in the parent model, before spontaneous supersymmetry breaking and finite temperature are switched on. There is a generic contribution to $n^u_T$ coming from the states arising generically in the moduli space of $S^1\times T^4/\Z_2$, while an additional one, $N_{\rm enhan}$, arises at special points of enhanced symmetry (see \cite{BEKP} and Sect. \ref{Rdyna}). $n^u_V=\sum_ {s=1}^{n^u_T}(-)^{\b Q_4(s)}$ depends on the choice of $\b Q_4$-charge operator. For the specific model considered at hand, one has
\be
\label{ntu}
\dis n_T^u=4\left[504+N_{\rm enhan}\right],\quad  \dis n_T^t=4\cdot 512\, , \quad n_V^u=4\left[248+\left((-)^{\b\eta}+(-)^{\b\eta'}\right)\, 128+N_{\rm enhan}\right].
\ee
We emphasize that the form of the free energy is the same for all choices of R-symmetry charge $\b Q_4$, with only the constants $n_T^u$ and $n_V^u$ changing.  The functions $f_T(z)$ and $f_V(z)$ arise from the relatively light Kaluza-Klein towers associated to the circles $S^1(R_0)$ and $S^1(R_4)$, and are given by,
\be
\label{fTV}
f_T(z)={\Gamma(5/2)\over \pi^{5/2}}\sum_{\tk_0,\tk_4}{e^{4z}\over \left[e^{2z}(2\tk_0+1)^2+(2\tk_4)^2\right]^{5/2}}\; , \qquad f_V(z)=e^{3z}\, f_T(-z)\, .
\ee

For comparison, the form of the free energy density for the background (I) as computed in \cite{BEKP} is (see appendix B),
\be
\label{F1}
{\rm (I)\; :} \qquad \F=-T^4\, \Big( (n^u_T+n_T^t)\, f_T(z)+(n^u_V+n^t_V)\, f_V(z)\Big) ,
\ee
where $n_{T}^u$, $n_T^t$, $n_V^u$ are given in Eq. (\ref{ntu}) and $n^t_V=\sum_ {s=1}^{n^t_T}(-)^{\b Q_4(s)}$ takes the value
\be
\label{ntv}
n_V^t=(-)^{\b\eta^{\prime\prime}}4\left(1+(-)^{\b\eta'}\right)256.
\ee


\subsection{$\N$ = 1 $\to$ 0 models: Case (III)}
\label{ZZ1}

We can extend the previous considerations to backgrounds of type (III), Eq. (\ref{backg3}). The first $\Z_2$ twists the directions $6,7,8,9$, while the second $\Z_2$ acts on the directions $4,5,6,7$. Again, $\N=1$ supersymmetry is spontaneously broken by the discrete deformation imposed by the non-trivial boundary conditions along the directions 0 and 4. Including the correct phases associated with the temperature and the spontaneous breaking of supersymmetry, the partition function is given by
\ba
\label{ZN=1}
\nonumber Z=&&\!\!\!\!\!\!\!\!\!\dis R_0R_{\rm box}^3\int_F{d\tau_1d\tau_2\over 2\tau_2^3}\dis{1\over 2}\sum_{H_1,G_1}{1\over 2}\sum_{H_2,G_2}{1\over 2}\sum_{a,b}(-)^{a+b+ab}{\theta[^a_b]\theta[^{a+H_1}_{b+G_1}]\theta[^{a+H_2}_{b+G_2}]\theta[^{a-H_1-H_2}_{b-G_1-G_2}]\over \eta^4}\\
\nonumber&&\dis \!\!\!\!\!\!\! \! \!{1\over 2}\sum_{\b\gamma,\b\delta}\b\theta[^{\b\gamma}_{\b\delta}]^8\, {1\over 2}\sum_{\b\gamma',\b\delta'}\b\theta[^{\b\gamma'}_{\b\delta'}]^5\b\theta[^{\b\gamma'+H_1}_{\b\delta'+G_1}]\b\theta[^{\b\gamma'+H_2}_{\b\delta'+G_2}]\b\theta[^{\b\gamma'-H_1-H_2}_{\b\delta'-G_1-G_2}]\, \sum_{n_0,\tm_0}e^{-{\pi R_0^2\over \tau_2}\abs \tm_0+n_0\tau\abs^2}(-)^{a\tm_0+bn_0+\tm_0n_0}\\
 &&\!\!\!\!\!\!\!\!\!\dis Z_{(1,1)}^{(a+\b Q_4,b+\b L_4)}[^{H_1}_{G_1}]\, Z_{(1,1)}^{(0,0)}[^{H_1}_{G_1}]\, Z_{(2,2)}^{(0,0)}[^{H_2}_{G_2}]\, Z_{(2,2)}^{(0,0)}[^{-H_1-H_2}_{-G_1-G_2}]\, {1\over \eta^2\bar\eta^{18}}\, ,
\ea
where the charges $\bar Q_4$, $\bar L_4$ are given in (\ref{charges}) and the blocks $Z_{(1,1)}^{(a+\b Q_4,b+\b L_4)}[^{H}_{G}]$ and $Z_{(n,n)}^{(0,0)}[^{H}_{G}]$ are given in (\ref{Z11}) and (\ref{Tn/Z2}).

- Clearly, the contribution of the $[^{H_2}_{G_2}]\equiv [^0_0]$ sector is half the result found for the background (I).

- For the sectors  $[^{H_2}_{G_2}]\not \equiv [^0_0]$, the arguments used for the background (II) apply identically. If $\b\eta=0$, the $\delta$-symbol in $Z_{(1,1)}^{(a+\b Q_4,b+\b L_4)}[^{H}_{G}]$ has no effect and can be forgotten, in which case the contributions are identical to the pure thermal case. If $\b\eta=1$, the only consequence of the $\delta$-symbol is to eliminate the subsectors $[^{\b\gamma}_{\b\delta}]\equiv [^{1-H_2}_{1-G_2}]$ from the answer found for the pure thermal case.

- For $[^{H_2}_{G_2}]\not \equiv [^0_0]$, the only subsectors that are not exponentially small (or vanishing) are $[^{H_1}_{G_1}\abs ^{H_2}_{G_2}]\equiv [^0_0\abs^{\, 1}_{G_2}]$ or  $[^{\, 1}_{G_2}\abs^{\, 1}_{G_2}]$, which happen to be equal. Thus, both imply contributions to the free energy density which are half the result found from the twisted sector of the background (II) in Eq. (\ref{F2}).

\noindent
Altogether, the pressure of the $\N_4=1\to 0$ model at finite temperature is
\be
\label{FN=1}
\F=-T^4\, \Big( (n^{u,u}_T+n_T^{t,u})\, f_T(z)+(n^{u,t}_T+n^{t,t}_T)\left(1-{\b\eta\over 2}\right) c_4+(n^{u,u}_V+n^{t,u}_V)\, f_V(z)\Big) ,
\ee
where the integer coefficients with index $T$ (or index $V$) count $+1$ (or $(-1)^{\bar Q_4}$) for each massless boson-fermion pair in the ((un)twisted,(un)twisted) sector of the $\Z_2\times \Z_2$ parent model (\ie before temperature and spontaneous supersymmetry breaking are turned on),
\be
\label{coef1}
\begin{array}{ll}
\dis n_T^{u,u}=2\left[504+N_{\rm enhan}\right], &  \dis n_T^{u,t}=2\cdot 512\, ,\\
\dis n_T^{t,u}=2\cdot 512\, , & \dis n_T^{t,t}=2\cdot 512\, ,\\
\dis n_V^{u,u}=2\left[248+\left((-)^{\b\eta}+(-)^{\b\eta'}\right)\, 128+N_{\rm enhan}\right]\, ,&
\dis n_V^{t,u}=(-)^{\b\eta^{\prime\prime}}2\left(1+(-)^{\b\eta'}\right) 256.
\end{array}
\ee

In the intermediate era, the above form of the free energy density naturally generalizes to a large number of heterotic models with $\N=1,2$ initial supersymmetry.
For any four-dimensional orbifold construction, with the restriction that the supersymmetry breaking flux wraps only one internal direction, the free energy density takes the general form:
\be
\label{F12}
\F=-T^4\, \Big( n_T\, f_T(z)+\tn_T\, c_4+n_V\, f_V(z)\Big) ,
\ee
where the integer coefficients satisfy
\be
\label{const}
n_T>0\; , \quad -n_T\le n_V\le n_T\; , \quad \tn_T\ge 0\, .
\ee
The origin of $\tn_T\, c_4$ is the contribution of the twisted sectors in which the internal  direction associated to the supersymmetry breaking is also twisted by the orbifold action. Indeed, in these sectors, the ``boson-fermion" mass splitting vanishes so that the contribution to the free energy is purely thermal.

When more internal cycles are wrapped with supersymmetry breaking fluxes, the free energy depends on $e^z=M/T$ and additional complex structure moduli.  For example, in the case when the supersymmetry breaking flux wraps two directions with radii $R_4$ and $R_5$, which are taken to be large compared to the string length in order to avoid Hagedorn-like divergences, the free energy yields a potential for the no-scale modulus $M = 1 / (2 \pi \sqrt{R_4 R_5})$ and the ratio $R_4/ R_5$ \cite{cosmo2}.  The general form of $\F$ involves a sum of integer contributions dressed by generalized functions $f_{T(V)}$.  In general, our approach can be applied to any string background based on a free field CFT such as free fermionic constructions, including the semi-realistic  Pati-Salam models of \cite{Assel:2010wj}.

We stress that (\ref{F12}) is valid only in the intermediate era, while at later times infra-red effects will become important.  These infra-red effects are highly model dependent, while our conclusions in the intermediate era are robust.  Namely, from this general form we shall find an attraction to a radiation-like era where the supersymmetry breaking scale is not constant but falls with a specific behavior determined by the expansion of the Universe, as well as a mechanism for stabilizing moduli.

In order to appreciate better the above form of the free energy, one has to compare it with that obtained via an effective field theory approach
(with initial $\N=1$ supersymmetry). Indeed, the result of the effective field theory approach ($EFT$) suffers from many ambiguities related to the UV cut-off scale $\Lambda_{co}$:
\be
-{\cal F}_{EFT}= T^4\,n^*\, c_4+  M^4 \,\left( C_4\, \ln {M\over T}+ \delta_4\right)\,+ C_2\left[ M^2\, \Lambda_{co}^2\, - C_4\,M^4\ln {\Lambda_{co}\over T}\right],
\ee
where $n^*$ is the effective number of light degrees of freedom below a given temperature $T$.  The term
$M^4\left( C_4\, \ln {M \over T}+ \delta_4 \right)$ is the re-normalized  effective potential,
(the renormalization point $\mu$ is taken here to be the temperature scale,  $\mu=T$). The  $M^4 \delta_4$ term re-normalizes the logarithmic divergences,  $M^4 C_4 \log {\Lambda_{co} \over T}$, while the quadratic divergences  $M^2 \Lambda_{co}^2$ are always  present  in any effective field theory with an initial supersymmetry $\N \le 2$, where $C_2$ is proportional to $Str{\cal M}^2$.
Thus, the field theory approach is unable to explain in simple terms the exact string result, even though it can be understood  why  $n^*\rightarrow n^*(z)$ becomes a function of $e^z=M/T$ once all Kaluza-Klein modes are taken into account (up to the scale $\Lambda_{co}$). In fact, the exact string computation used here and in previous works \cite{cosmo1,cosmo2,BEKP} is necessary to take into account the gravitational, the observable and the hidden sector contributions to $\F$ (they all contribute to the integer coefficients such as $n_T$, $n_V$, $\tilde n_T$, etc...) and to justify the absence of any $M^2 M_{String}^2$ term.


\section{Attraction to radiation-like dominated eras}
\label{radera}

In the previous section, we first computed the free energy density $\F$ for the simple backgrounds (I), (II), (III) defined in Eqs (\ref{backg1})--(\ref{backg3}), when the Universe evolves in the intermediate era.  Supersymmetry breaking was implemented by a single geometrical flux in the internal direction 4.  In (\ref{F12}) we gave the form of the free energy density for general orbifold models where supersymmetry is broken by geometrical flux in a single internal direction.  The moduli associated to the directions $5,\dots,9$ are assumed to generate scales much larger than $T$ and $M$, as will be justified in Sect. \ref{Rdyna}.  In this section we proceed to analyze the cosmology induced by the general form of the free energy density (\ref{F12}).  We will find an attraction to a radiation-like evolution, where the quantity $e^z = M/T$ can be stabilized.  This provides a mechanism for generating the hierarchy between the string scale and the supersymmetry breaking scale.  This result is robust and remains valid for the semi-realistic Pati-Salam models of \cite{Assel:2010wj}.

The behavior naturally splits into two types: the first one is illustrated by case (I), (\ref{backg1}), and corresponds to having $\tilde n_T=0$ in the general form given in (\ref{F12}).  The second one is illustrated by cases (II) and (III), defined in (\ref{backg2}) and (\ref{backg3}), and corresponds to having $\tilde n_T > 0$. For case (I), the study of the back-reaction of the free energy density on the initially static background was initiated in \cite{Bourliot:2009na}.  For cases (II) and (III), an important role will be played by the additional contribution from the twisted sectors, the net result of which is to strengthen the attraction to the radiation-like evolution.

We are interested in isotropic and homogeneous cosmologies. More specifically, we take the dilaton $\phi$ as well as $R_4$ to depend only on time, while the space-time metric is assumed to be of the form $ds^2 = - dt^2 + a(t)^2 \left[ (dx^1)^2 + (dx^2)^2 + (dx^3)^2 \right]$.   The remaining scalars are fixed to constant values, while the gauge fields are taken to be pure gauge.  In Einstein frame, the four-dimensional low energy effective action takes the form
\be
\label{S}
S=\int d^4x \sqrt{-g} \left[{R\over 2}-{1\over 2}(\partial \Phi)^2-{1\over 2}(\partial \phi_\bot)^2+\cdots-\F\right]\, ,
\ee
where we have introduced the following notations
\ba
\label{champs}&\dis\Phi:= \sqrt{2\over 3}\, (\phi-\ln R_4)\; , \qquad \phi_\bot:={1\over \sqrt{3}}\, (2\phi+\ln R_4)\; ,&\\
\label{def4} &\F=-T^4\, \dis p(z) \; , \quad e^z:={M\over T}\equiv {R_0\over R_4}\; ,\quad M={e^{\sqrt{3\over 2}\Phi}\over 2\pi}={1\over 2\pi R_4\, e^{-\phi}}\; , \quad T={1\over 2\pi R_0\, e^{-\phi}}.&
\ea
Note that in (\ref{S}), the free energy density $\F$ is only a non-trivial source for the four-dimensional metric and $\Phi$ (or $M$) which involves the dilaton $\phi$ and scalar $\ln R_4$. For our FLRW metric, the contribution of the free energy density to the stress-energy tensor takes the form ${T_\mu}^\nu={\rm diag}{(-\rho,P,P,P)_\mu}^\nu$, where we have introduced the energy density $\rho$ and the pressure $P$. They are found using the variational principle only,
\be
\label{pr}
P=T^4\, p(z)\; ,\qquad \rho=T\, {\partial P\over \partial T}-P:= T^4\, r(z)\qquad \where\qquad  r(z)=3p-p_z\, ,
\ee
and reproduce standard expressions derived from the axioms of thermodynamics.
The equations of motion and the expression of the conservation of the stress-energy tensor can be found in \cite{Bourliot:2009na}. They are more conveniently written in terms of $(\ln a)$-derivatives, using $\dis\dot y=H{dy\over d\ln a}:=H\o y$ (for any field $y(t)$). The conservation of the stress-energy tensor yields
\be
\label{Ta}
T={e^{\A(z)-z}\over a}\qquad \with\qquad \A_z(z)={4r-r_z\over 3(r+p)},
\ee
while the Friedmann equation takes the form
\be
\label{Fried}
H^2=T^4\, {r(z) \over 3-\K(z,\overset{\circ}{z},\overset{\circ}{\phi}_\bot)}\qquad \where
\qquad \K={1\over 3}\left(\A_z(z)\overset{\circ}{z}-1\right)^2+{1\over 2}\, \overset{\circ}{\phi}_\bot\!\!\!{}^2\, .
\ee
For the scalar fields, it is more appropriate to consider equations for $z$, defined in (\ref{def4}),  (rather than $\Phi$) and $\phi_\bot$,\footnote{Note that we are in the early history of the universe and so the decay rates of the moduli $\Phi$ and $\phi_\perp$ are Planck suppressed and negligible compared to the gravitational friction coming from the expansion of the Universe.}
\be
\label{syseq}
\left\{
\begin{array}{l}
\displaystyle
{r(z)\over 3-\K(z,\overset{\circ}{z},\overset{\circ}{\phi}_\bot)}\left(\A_z(z)\,
  \overset{\circ\circ}{z}+\A_{zz}(z)\, \overset{\circ}z
  \!\!\!\phantom{x}^2\right)+{r(z)-p(z)\over 2}\,  \A_z(z)\,
\overset{\circ}{z}+V_z(z)=0\\ \\
\displaystyle {r(z)\over 3-\K(z,\overset{\circ}{z},\overset{\circ}{\phi}_\bot)}\,
\overset{\circ\circ}{\phi}_\bot +{r(z)-p(z)\over 2}\,
\overset{\circ}{\phi}_\bot =0,
\end{array}
\right.
\ee
where $V(z)$ is defined by its $z$-derivative,
\be
\label{r4p}
V_z(z)=r(z)-4p(z)\, .
\ee
Note that the auxiliary potential $V(z)$ should not be confused with the thermal effective potential $- {\cal F}$.

The system of equations (\ref{syseq}) is highly non-linear and allows drastically different behaviors, depending on the IBC. However, a particularly interesting one can occur when $V(z)$ admits a critical point, $V_z(z_c)=0$ \cite{Bourliot:2009na,Bourliot:2009cx,Antoniadis:1986ke}. When such an extremum exists, the equations (\ref{syseq}) admit the particular solution $z\equiv z_c$, $\phi_\bot\equiv cst$. From (\ref{Ta}) and (\ref{Fried}), this evolution describes effectively a radiation-like era in four dimensions similar to the one quoted in the introduction, (\ref{RDS}).  This can be seen by noting we have
\be
\label{rad4}
3H^2= {C_r\over a^4}\qquad \where \qquad C_r={9\over 2}\, p(z_c)\, e^{4[\A(z_z)-z_c]}\, ,
\ee
that results from the proportionality of $M(t)$, $T(t)$ and $1/a(t)$,
\be
\label{prop}
M(t)=T(t)\times e^{z_c}={1\over a(t)}\times e^{\A(z_c)}\quad \with \quad a(t)=\sqrt{t}\, \left({4\over 3} C_r\right)^{1/4} ,\quad \phi_\bot(t)=cst.
\ee
Of course, a subsidiary condition for the solution (\ref{rad4}) to exist is $p(z_c)>0$, as checked below. Note that geometrically, this solution satisfies
\be
\label{propo}
R_0(t)\propto R_{\rm box}(t)\propto R_4(t)\, ,
\ee
while the dilaton decreases, as follows from the definition of $\phi_\bot$ in Eq. (\ref{champs}).
Moreover, even if the evolution (\ref{rad4})--(\ref{prop}) mimics a radiation dominated Universe, {\em it is not radiation dominated}. This is clear from Eq. (\ref{r4p}) that is telling us that the state equation of the thermal bath of string states is
\be
\label{cri}
\rho=4 P\, .
\ee
Since the mass of the KK states in the direction 4  decreases as the temperature, the massive KK states never decouple and the thermal system is never dominated by radiation.
As announced in the introduction, it is only by adding the contributions $\rho_{\rm kine}=P_{\rm kine}=\dis { 1\over 2}\dot \Phi^2$ to their thermal counterparts that the state equation of the {\em total system} thermal + kinetic is radiation-like, $\rho_{\rm tot}=3 P_{\rm tot}$ \cite{cosmo1,cosmo2,BEKP}.

\begin{figure}[h!]
\begin{center}
\vspace{.3cm}
\includegraphics[height=3.5cm]{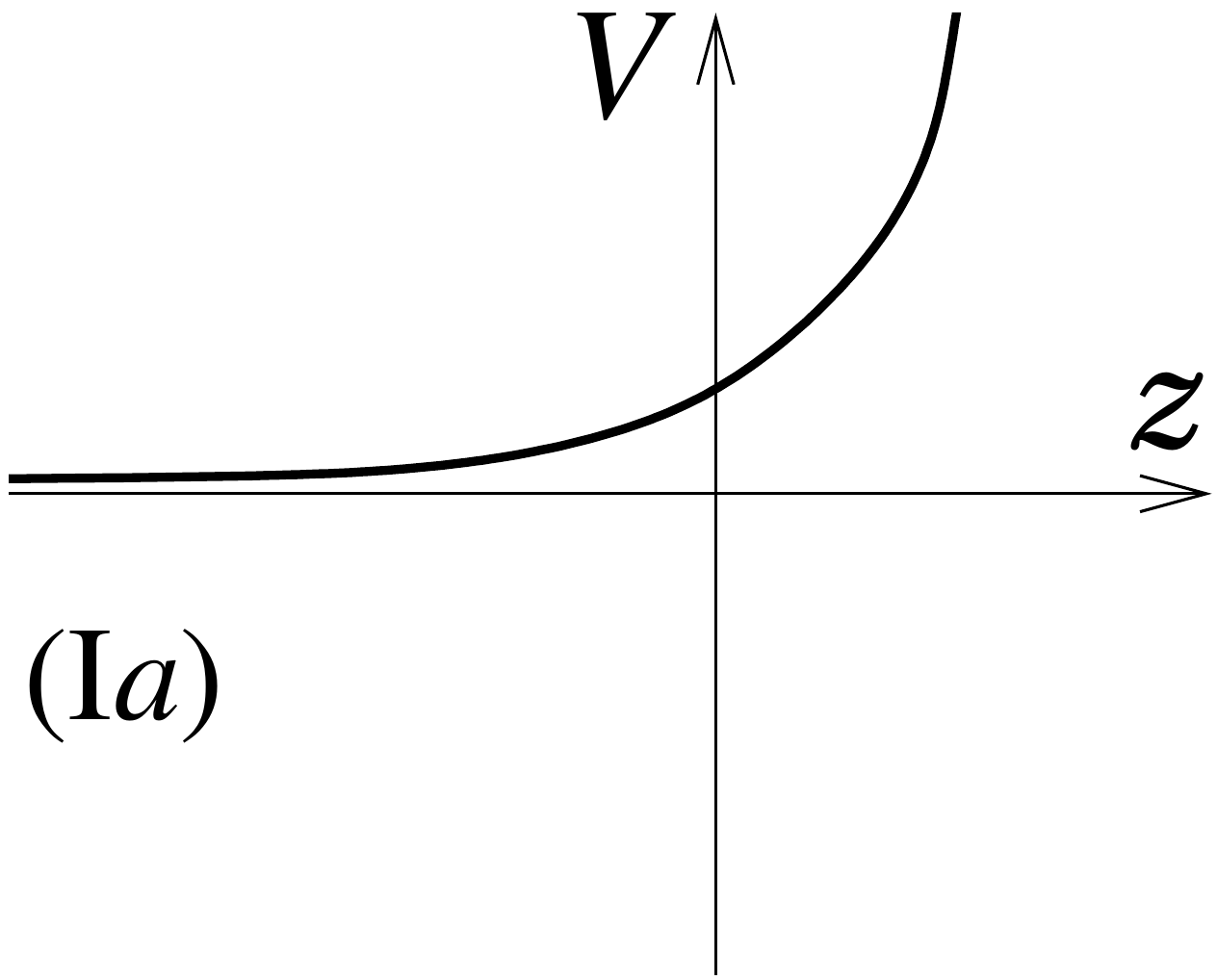}\qquad\quad
\includegraphics[height=3.5cm]{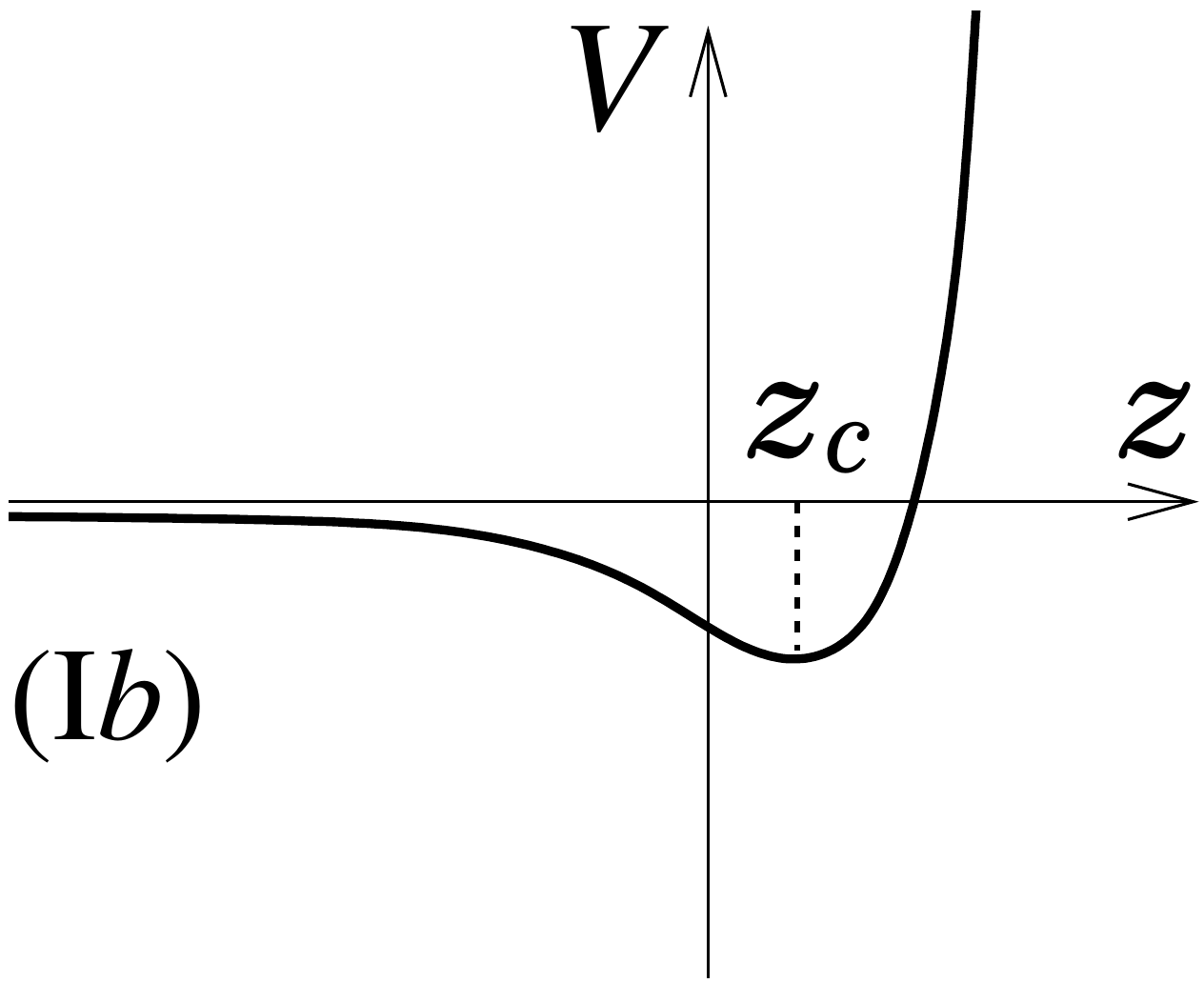}\qquad\quad
\includegraphics[height=3.5cm]{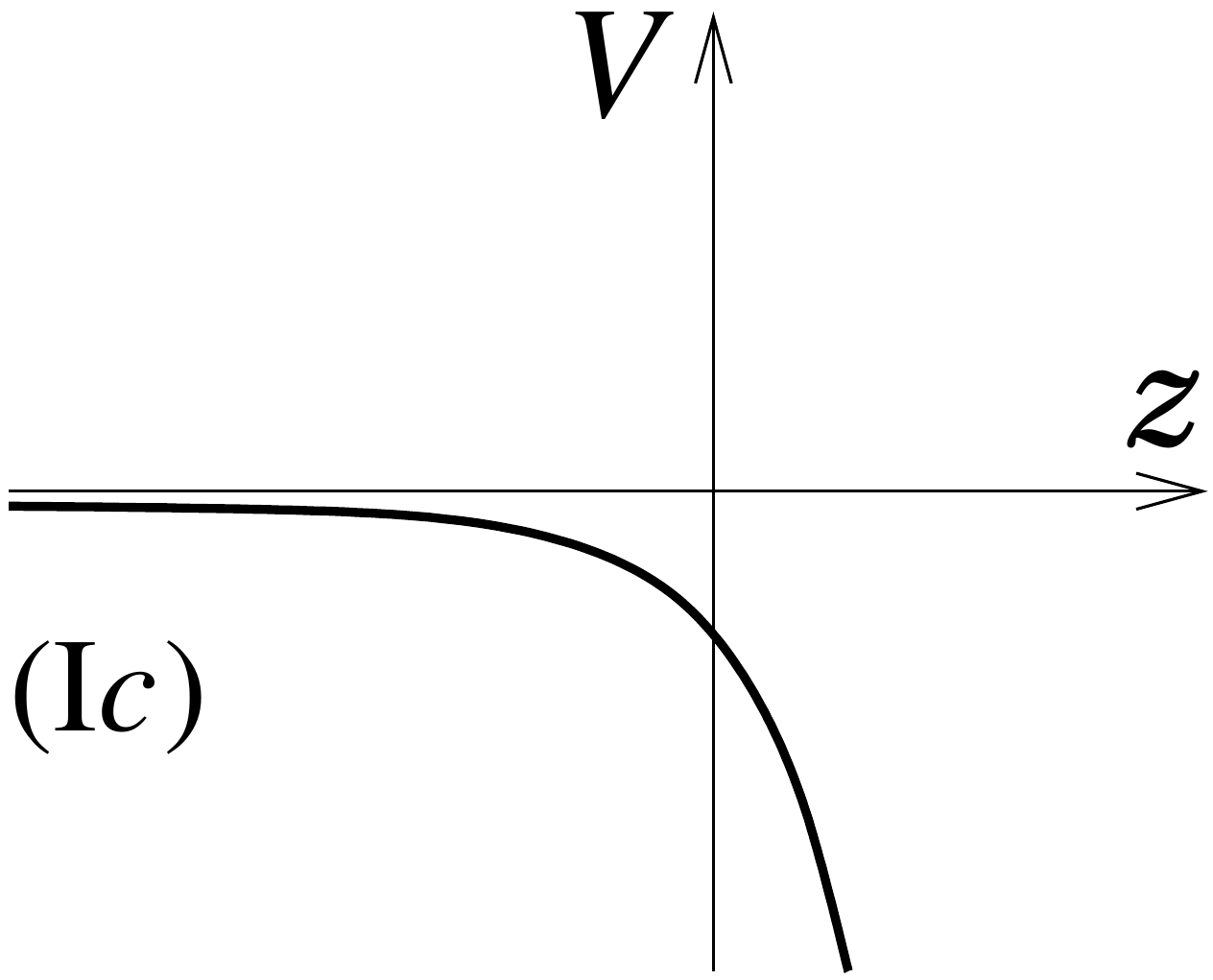}\\
\vspace{.4cm}
\includegraphics[height=3.5cm]{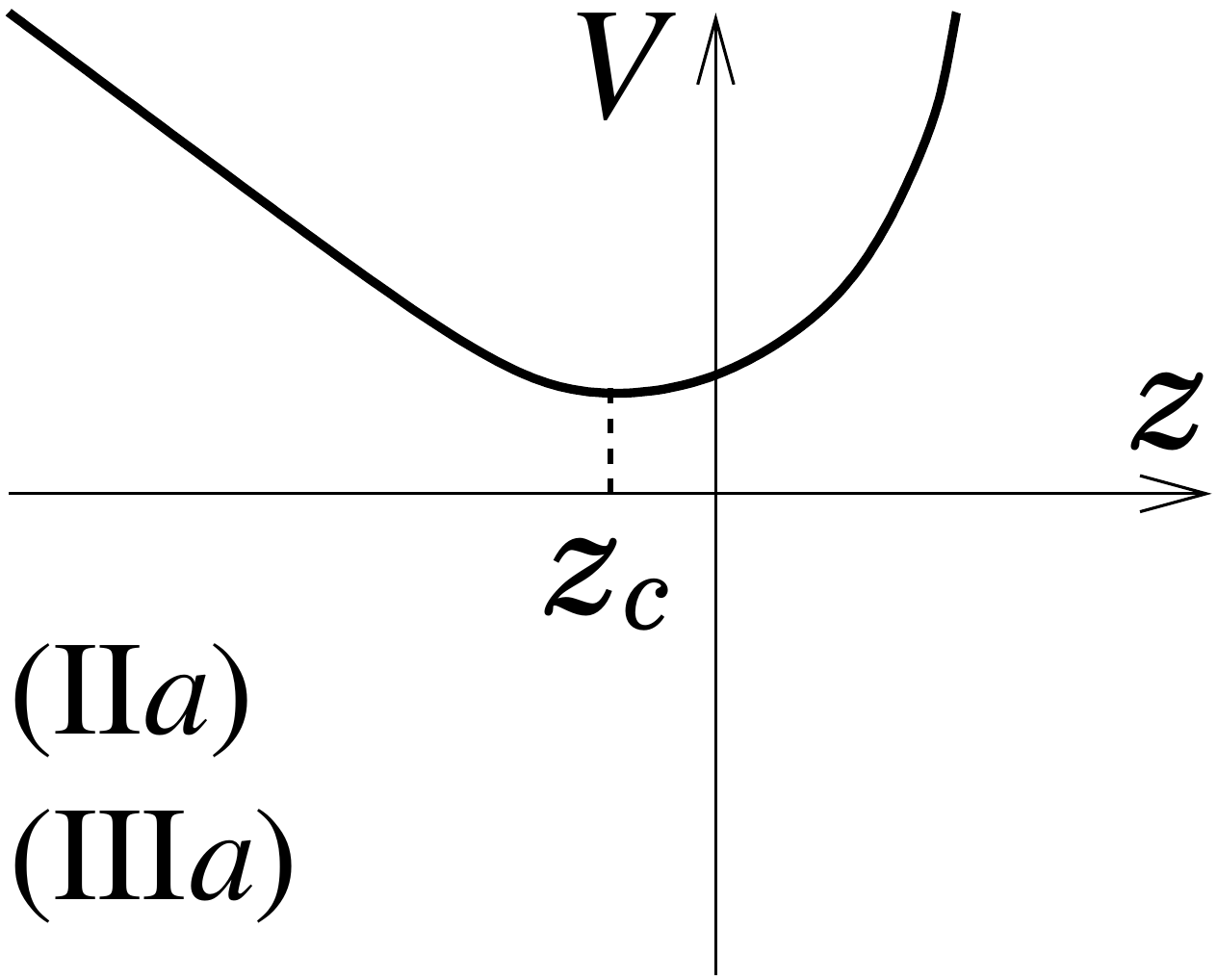}\qquad \quad
\includegraphics[height=3.5cm]{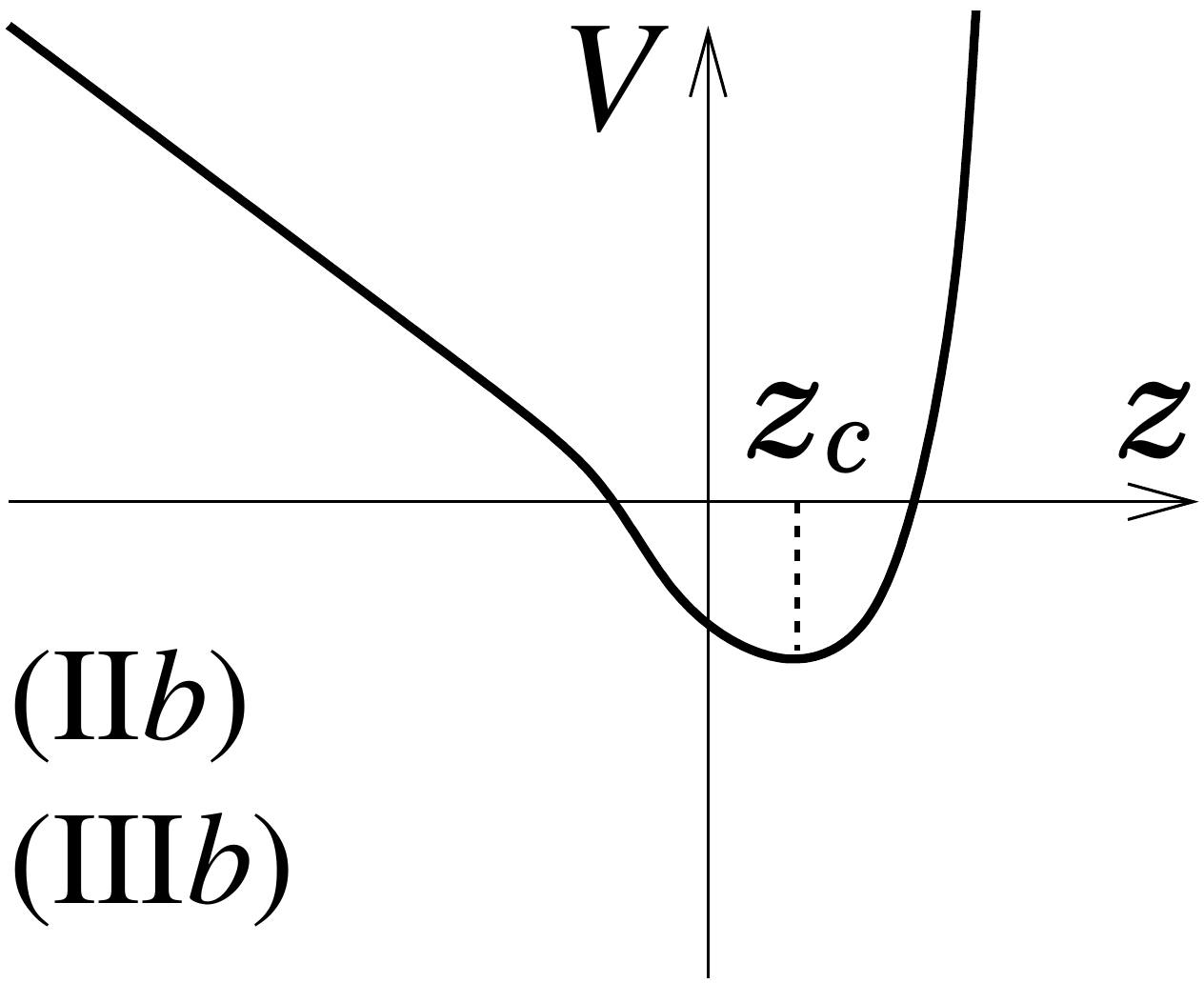}\qquad\quad
\includegraphics[height=3.5cm]{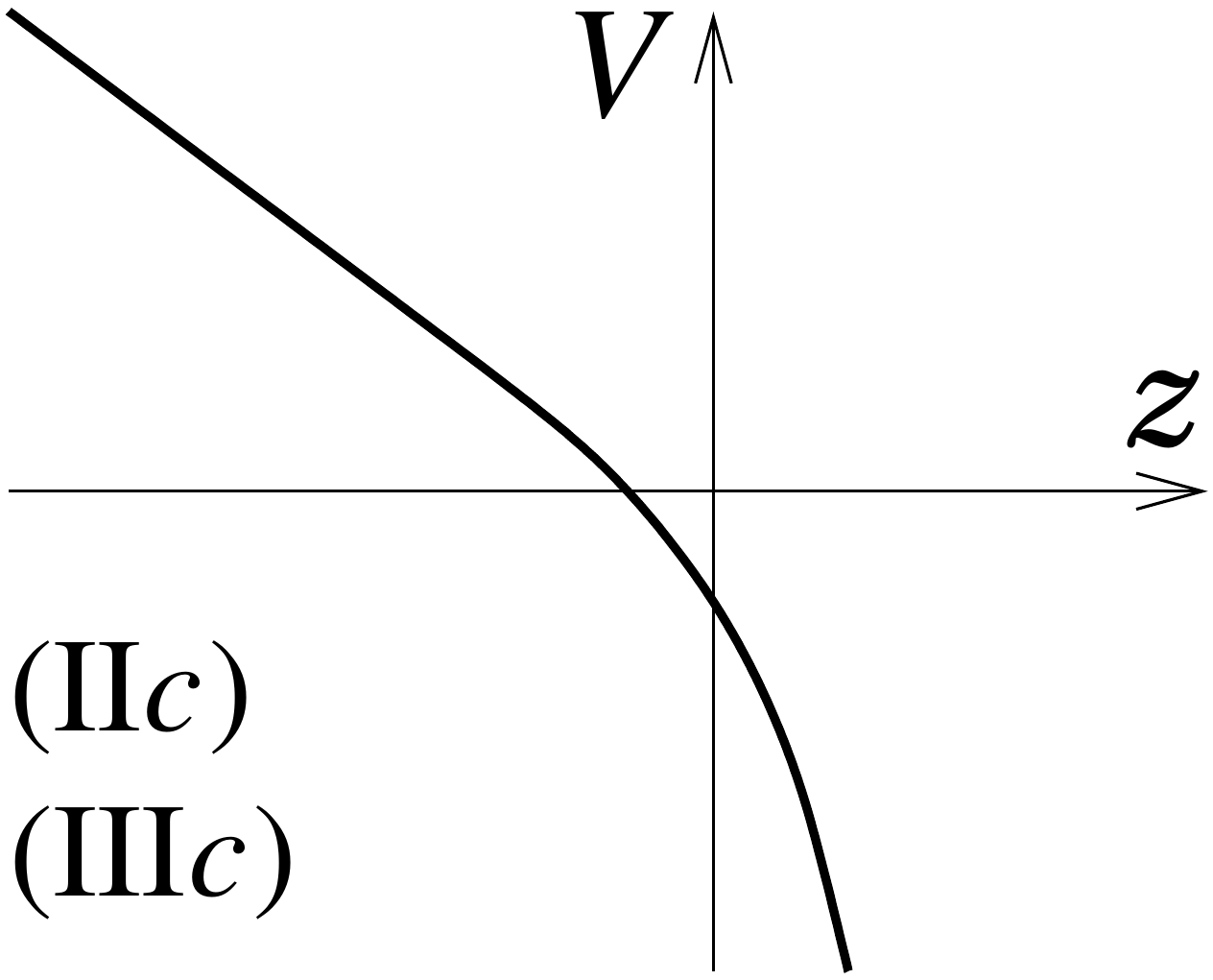}
\caption{\footnotesize \em The qualitative behavior of $V(z)$ defined in (\ref{r4p}) depends on the value of the parameter $n_V/n_T$. The ranges $(a)$, $(b)$ and $(c)$ correspond to $-1< n_V/n_T\le-1/15$, $-1/15<n_V/n_T<0$ and $0<n_V/n_T\le 1$, respectively. For a pressure in case {\rm (I)}, an extremum exists only in the range $(b)$. For a pressure in case  {\rm (II)} or {\rm (III)}, an extremum occurs in the ranges $(a)$ and $(b)$ i.e. $n_V<0$. (Drawing is with arbitrary vertical origin.)}
\vspace{-.4cm}
\label{fig_V}
\end{center}
\end{figure}

We now turn to the problem of  determining when a critical point $z_c$ exists, in order  to classify the models. The shape of the auxiliary potential $V(z)$ was studied in case (I) in \cite{Bourliot:2009na}. For a given value of the model-dependent ratio  $\dis{n_V/ n_T}$, three different behaviors are allowed\footnote{The case $n_V=0$ corresponds to a pure thermal case and is analyzed in appendix A.},
\be
\label{Iabc}
\begin{array}{lll}
\mbox{(I$a$)}&:\dis \qquad  {n_V\over n_T}\le -{1\over 15}&, \quad\mbox{no extremum,}\\
\mbox{(I$b$)}&:\dis \qquad -{1\over 15}<{n_V\over n_T}<0& , \quad\mbox{one extremum,}\\
\mbox{(I$c$)}&:\dis \qquad 0<{n_V\over n_T}&, \quad\mbox{no extremum.}
\end{array}
\ee
In case (II) and (III), $V(z)$ gets an additional linear  contribution, $\dis -z\times \tn_T c_4$, that unifies the above ranges $(a)$ and $(b)$ and will play an important role in the next section,
\be
\label{IIabc}
\begin{array}{lllll}
\mbox{(II$a$), (II$b$)}& \mbox{and}& \mbox{(III$a$), (III$b$)}&:\dis \qquad n_V<0& , \quad\mbox{one extremum,}\\
\mbox{(II$c$)} &\mbox{and}& \mbox{(III$c$)}&:\dis \qquad 0<n_V&, \quad\mbox{no extremum.}
\end{array}
\ee
Fig. \ref{fig_V} shows the qualitative shapes of the auxiliary potential $V(z)$ in all cases.
Some remarks are in order:

$\bullet$ In the range ($a$) of the parameter $\dis{n_V/ n_T}$, we see that no matter how small $\tn_T>0$ may be for the backgrounds (II) and (III), its effect is drastic since it lifts the asymptotic flat direction of $V(z)$ present in case (I).

$\bullet$ In all instances where an extremum exists, the sign of $p(z_c)$ is positive and the radiation-like era (\ref{prop}) is valid.
The stability of this solution is guaranteed by showing that small time-dependent perturbations always die off.  In particular, we find that $\overset{\circ}{z} \rightarrow 0$ as $a \rightarrow \infty$, showing that $\dot z = \o z H\ll H$. From this we conclude that the contribution of the kinetic energy of $z$ to $H^2$ is negligible.  This is sufficient to show that at late times, any energy stored in the oscillations of the modulus $z$ dilutes faster than the thermal energy.  This was done explicitly in \cite{Bourliot:2009na} for  pressures in case (I) and remains true in cases (II) and (III). Thus,  the solution $z\equiv z_c$, $\phi_\bot\equiv cst$ is an attractor, and we conclude that for generic IBC, the solution $z(t)$, $\phi_\bot(t)$ converges to it. The evolution (\ref{prop}) is thus a critical solution. In a neighborhood of $(z,\o z,\o \phi_\bot)\simeq (z_c, 0,0)$, the linearized equation of $z$ in (\ref{syseq}) shows that the system behaves as a damped oscillator, so that $V(z)$ can be interpreted as a potential in this regime, as suggested by Fig. \ref{fig_V}.

$\bullet$ For completeness, we signal that in the range ($c$) of the parameter $\dis{n_V/ n_T}$ \ie $n_V>0$, the analysis \cite{Bourliot:2009na} in case (I) applies identically to the cases (II) and (III). The Universe can be attracted to a ``Moduli Dominated Era'' of contraction. However, the range of validity of this behavior is restricted by the fact that the perturbation of the classically static backgrounds increases with time.


\section{Dynamical change of space-time dimension}
\label{T>M}

In this section, our aim is to determine the induced cosmology in case (I$a$). For such a background,   the shape of $V(z)$ in Fig. \ref{fig_V} suggests that $z(t)$, instead of being attracted and stabilized to some critical value $z_c$, may reach large negative values and behave as a modulus along a flat direction. In such a regime $T \gg M$, the statistical effects are expected to screen the pure quantum effects (present at zero temperature).  In fact, for small $\dis e^z={R_0/ R_4}$, the free energy density ({\ref{F12}) for all backgrounds (I), (II) or (III) becomes
\be
\label{z<-1}
\F=-{T^5\over M}\left(n_T\,  c_5+\left({R_0\over R_4}\right) \tn_T\, c_4+\left({R_0\over R_4}\right)^4\left({n_T\over 15}+n_V\right){c_4\over 2}+\cdots\right) ,
\ee
where $c_{4,5}$ are defined in Eq. (\ref{Zstef}) and the neglected terms are of order $\O(e^{-2\pi(R_0/R_4)^{-1}})$.
We are going to see that this expression is more naturally interpreted from a five-dimensional point of view, as suggested by the large hierarchy $R_4\gg R_0$.

To avoid confusions with fields normalized in four dimensions, we denote all quantities in five dimensions with primes. In particular, we reinterpret $R_4$ as the radius of a fourth external spatial direction. In total, the Universe is five-dimensional, homogeneous but anisotropic, with Einstein frame scale factor $a'$ in the directions 1,2,3 and $b$ in the direction 4. Together with the dilaton $\phi':= \sqrt{3}\, \phi'_\bot/2$ and the temperature $T'$, their definitions can be found in Eq. (\ref{ansatz5}). Instead of $b$,  it is convenient to work with the field
\be
e^\xi:= {b\over a'}\, .
\ee
We derive the equations of motion  in appendix D. The conservation of the stress-energy  tensor can be found in Eq. (\ref{e2'}), the Friedmann equation in (\ref{e1'}) and  the equations for $\xi$ and $\phi_\bot'$ in (\ref{e3'}) and (\ref{e4'}).  As in Sect. \ref{radera}, they are written  in terms of $(\ln a')$-derivatives.

We solve the equations in two steps. First, we suppose $\dis {R_0/ R_4}$ small and neglect the linear and quartic monomials in the partition function or free energy density, Eq. (\ref{z<-1}).  Under this approximation,  the thermal gas is found to satisfy the state equation for radiation in five dimensions, as follows from Eq. (\ref{source}),
\be
\label{st5}
\rho'=4P'=4\, T^{\prime 5}\, n_T\, c_5.
\ee
Correspondingly, the four equations of motion admit the particular solution where $(z,\xi,\phi_\bot')$ are all constants, $(z_0,\xi_0,\phi_{\bot 0}')$, while the metric and temperature evolutions are that of a radiation dominated Universe in five dimensions,
\be
\label{att}
b(t)=a'(t)\times e^{\xi_0}={1\over T'(t)}\times e^{-z_0}\qquad \where \qquad 6H^{\prime 2}={C_r'\over a^{\prime 5}}\; ,\quad C_r'=4n_Tc_5e^{-5(z_0+\xi_0)}.
\ee
In appendix E, we show that for arbitrary IBC, $\o \xi$, $\o \phi{}_\bot'$ (and thus $\o z$) are converging to 0 at late times.  This means that the solution (\ref{att}) is an attractor whenever we may neglect the subdominant terms in (\ref{z<-1}).

Second,  we study the effect on the dynamics of the ``residual forces'' that arise from the subdominant terms $e^z$ or $e^{4z}$ we have neglected in Eq. (\ref{z<-1}). The conclusions can be inferred from the shapes of the potentials $V(z)$ in four dimensions, shown in Fig. \ref{fig_V}.  In case (I$a$), $V(z)$ is slightly increasing when $e^z \ll 1$, so that the residual force is expected to push $z$ towards more negative values. However, the conclusions look the opposite in all other cases.
To be quantitative, we consider a perturbation around the solution (\ref{att}),
\be
\label{pert}
z=z_0+\varepsilon_{(z)}\;, \qquad  \xi=\xi_0+\varepsilon_{(\xi)}\;, \qquad \phi'_\bot =\phi'_{\bot 0}+\varepsilon_{(\phi'_\bot)}\; ,
\ee
where $\abs\varepsilon_{(z)}\abs$, $\abs\varepsilon_{(\xi)}\abs$, $\abs\varepsilon_{(\phi'_\bot)}\abs$ and $e^{4z_0}$ in case (I) ($e^{z_0}$ in case (II) and (III)) are $\ll 1$. At order one, Eqs (\ref{e2'}), (\ref{e3'}) and (\ref{e4'}) become
\be
\label{epsi}
\o\varepsilon_{(z)}=-{3\over 4}\, \o \varepsilon_{(\xi)}\; , \qquad \oo \varepsilon_{(\xi)}+{3\over 2} \left(\o\varepsilon_{(\xi)}+C\right)=0\; ,\qquad  \oo \varepsilon_{(\phi'_\bot)}+{3\over 2}\, \o\varepsilon_{(\phi'_\bot)}=0\; ,
\ee
where the definition of the constant $C$ differs if $\tn_T=0$ or $\tn_T>0$,
\be
\label{C}
C=e^{4z_0}\, 2\, {c_4\over c_5}\left({n_V\over n_T}+{1\over 15}\right)\mbox{ in case (I)} \; , \qquad C=e^{z_0}{c_4\over c_5}{\tilde n_T\over n_T}\mbox{ in case (II) and (III)}\, .
\ee
The solutions to Eqs (\ref{epsi}) are
\be
\label{conv}
\o\varepsilon_{(z)}= {3\over 4}\, C\; , \qquad \o\varepsilon_{(\xi)}= -C\; , \qquad \o\varepsilon_{(\phi'_\bot)}=0\; ,
\ee
and some consequences follow:

$\bullet$ In case (I$a$), we have $C\le 0$ and, as expected, the residual force (compensated by the friction due to the expansion of the universe) induces a small negative velocity for $z$ that decreases. The perturbation (\ref{pert}) is thus more and more negligible and the attraction to the solution (\ref{att}) is justified.

$\bullet$ In all other backgrounds, (I$b$), (I$c$) and cases (II) and (III), we have $C>0$. The residual force implies that $z$ increases and the approximation $e^z\ll 1$ breaks down at some point. This is consistent with the fact that the system is actually going back to a regime $T\not \gg M$ and converges to one of the attractors found in Sect. \ref{radera} in four dimensions.

$\bullet$ In the latter situation, the force that is pulling the system out of the regime $T\gg M$ is much stronger in cases (II) and (III), as compared to cases (I$b$) and (I$c$). This is clear from the definition of $C$ in Eq. (\ref{C}) that is linear in $\dis {R_0/ R_4}$ in case (I) and quartic in case (II) and (III), or from the slopes of the potentials in Fig. \ref{fig_V} for $e^z\ll 1$.

Our conclusion is that in all cases where $n_V<0$, the cosmological evolutions are attracted to radiation(-like) solutions. Geometrically, they satisfy the proportionality relation (\ref{propo}). However, important remarks need to be pointed out. In case (I$a$),  $z$ is not stabilized but only frozen at late times at a value which is not uniquely determined by $\dis {n_V/ n_T}$, \ie the properties of the massless spectrum. In that sense, $z$ enters dynamically into a phase where it behaves as a modulus. Its final value $z_0$ depends on additional data, namely the IBC at the exit of the Hagedorn era, and in particular the initial kinetic energies. Thus, the backgrounds (I$a$) provide a natural explanation for the dynamical appearance of a hierarchy $R_4\gg R_0$ which justifies that the radius $R_4$ should not be treated as an internal space radius (that determines a supersymmetry breaking scale screened by thermal effects). Instead, the backgrounds (I$a$) yield a dynamical decompactification of the direction 4 and stabilize to five the space-time dimension.  In the other cases where $n_V<0$, we may also attempt to introduce this hierarchy by adjusting the value of $z_c$; however, it is only at the price of a very artificial choice of the model that the critical value $z_c$ would satisfy $e^{z_c} \ll 1$.


\section{Internal space dynamics}
\label{Rdyna}

In the previous sections, we have concentrated our attention on the dynamics of the moduli which determine the scales of spontaneous supersymmetry breaking, namely $M$ and $T$, under the assumption that the other scales characterized by internal moduli were much larger than $M$ and $T$. For instance, the internal radii were taken to satisfy the inequalities (\ref{hyp}). Under this hypothesis, we have shown that after the Hagedorn era, the 1-loop free energy density (or thermal effective potential) is independent of them, up to exponentially suppressed terms (see Eq. (\ref{moduli})). As a result, we were allowed to keep these moduli frozen and look for the cosmological evolutions of the remaining degrees of freedom. Our aim in this section is to argue that the above properties of the internal moduli are actually not restrictions but inevitable consequences of attraction phenomena. In particular, the constraint (\ref{hyp}) will be satisfied at late enough times, for arbitrary IBC at the exit of the Hagedorn era.

We suppose the constraint (\ref{hyp}) is valid for all radii $R_I$ except $R_6$ (or $R_5$), which is kept arbitrary. Thus, the dependance of the free energy density with respect to $R_6$ (or $R_5$) is in general no longer exponentially small. This implies that we need to examine cosmological evolutions with non-trivial dynamics for this radius.  Depending on the initial data for the radius, we shall find that it either behaves as a modulus or is stabilized at its self-dual point with a mass scale of the order $T \sim M$.  For large enough initial values, the internal cycle may experience a period of expansion, after sufficient time this expansion halts and the evolution settles into the RDS$^4$ defined in (\ref{RDS}).\footnote{The question of the stabilization of the internal space in the simpler case where supersymmetry is spontaneously broken by thermal effects only is considered in appendix A with similar results.}

\subsection{$\N$ = 2 $\to$ 0 models: Cases (I) and (II) \label{sectn2to0nonsusydyn}}

We have considered in Sect. \ref{FZ2} backgrounds where the direction 4, associated with supersymmetry breaking, is either a circle (case (I)) or part of a $\Z_2$-orbifold (case (II)), see (\ref{backg1}) and (\ref{backg2}) respectively. Similarly, we can take the internal direction 6 to be a circle or part of the orbifold. We thus have four cases to analyze:
\be
\label{models}
\begin{array}{lll}
\mbox{case (I.i)}&: & S^1(R_0)\times T^{3}(R_{\rm box})\times S^1(R_4)\times S^1(R_6) \times \dis {T^4 \over \Z_2}\\
\mbox{case (I.ii)} &: &S^1(R_0)\times T^{3}(R_{\rm box})\times S^1(R_4)\times S^1 \times \dis {S^1(R_6) \times T^3 \over \Z_2}\\
\mbox{case (II.i)} & :&S^1(R_0)\times T^{3}(R_{\rm box})\times S^1(R_6)\times S^1 \times \dis {S^1(R_4) \times T^3 \over \Z_2}\\
\mbox{case (II.ii)} & :&S^1(R_0)\times T^{3}(R_{\rm box})\times T^2 \times \dis {S^1(R_4) \times S^1(R_6) \times T^2 \over \Z_2}.
\end{array}
\ee
To write the free energy density, we separate the untwisted and twisted sector's contributions,
$\F = \F^u + \F^t$. The first part, $\F^u$, is common to the four classes of models and is computed in \cite{BEKP}. Its explicit form,
\be
\label{fu}
\begin{array}{ll}
\F^u(z,\eta,\zeta) = - T^4 \, \Big( &\!\!\!\!n_T^u \, \big( f_T(z) + k_T(z,\eta - |\zeta|)\big) + n_V^u\,  \big(f_V(z) + k_V(z,\eta - |\zeta|)\big)\\
&\!\!\!\! + \, \hat n_T^u \, g_T(z,\eta,|\zeta|) + \hat n_V^u \, g_V(z,\eta,|\zeta|) \Big)
\end{array}
\ee
is expressed in terms of the scalar fields
\be
e^z={M\over T}={R_0\over R_4}\; ,~~ \qquad e^\eta = R_4\; , ~~\qquad e^\zeta = R_6\, .
\ee
$f_T$ and $f_V$ are defined in Eq. (\ref{fTV}), while the remaining functions are
\be
\begin{array}{l}
k_T(z,\eta-\abs\zeta\abs)=\displaystyle {\sum_{m_6}}'\abs m_6\abs^{5/2}e^{{5\over 2}(\eta-\abs\zeta\abs)}e^{4z} \sum_{\tk_0,\tk_4}{2K_{5/ 2}\left(2\pi\abs m_6\abs e^{\eta-\abs\zeta\abs}\sqrt{e^{2z}(2\tk_0+1)^2+(2\tk_4)^2}\right)\over \left[e^{2z}(2\tk_0+1)^2+(2\tk_4)^2\right]^{5/ 4}}\\
k_V(z,\eta-\abs\zeta\abs)=e^{3z}~k_T(-z,\eta-\abs\zeta\abs+z)
\end{array}
\ee
\be
\begin{array}{l}
g_T(z,\eta,\abs\zeta\abs)=\displaystyle \left(e^{2\abs\zeta\abs}-1\right)^{5/ 2} e^{{5\over 2}(\eta-\abs\zeta\abs)}e^{4z}\sum_{\tk_0,\tk_4}{2K_{5/ 2}\left(2\pi (e^{2\abs\zeta\abs}-1)e^{\eta-\abs\zeta\abs}\sqrt{e^{2z}(2\tk_0+1)^2+(2\tk_4)^2}\right)\over \left[e^{2z}(2\tk_0+1)^2+(2\tk_4)^2\right]^{5/4}}\\
g_V(z,\eta,\abs\zeta\abs)=e^{3z}~g_T(-z,\eta+z,\abs\zeta\abs)\, ,
\end{array}
\ee
where $K_{5/2}$ are modified Bessel functions of the second kind.
In $\F^u$, the coefficients $n_T^u$ and $n_V^u$ are defined in Eq. (\ref{ntu}). They count the massless boson-fermion pairs (in the parent supersymmetric model) for generic values of $R_6$. However,  the self-dual point $R_6=1$ is special, due to an $SU(2)$ enhancement of the gauge symmetry. The additional massless states at this point are taken into account by the terms in the second line of Eq. (\ref{fu}), with
\be
\hat n^u_T=\hat n^u_V=4\cdot 2\, .
\ee
The role of the functions $g_T$ and $g_V$ is to interpolate between the generic and extended spectrums. On the contrary, the role of $k_T$ and $k_V$ is to interpolate between a four dimensional point of view (when $R_6$ is close to one) and a five dimensional  one (when $R_6$ or $1/R_6$ is very large).

We now move onto describing the twisted contributions $\F^t$ for each background:

- Case (I.i) : The result for these models is derived in \cite{BEKP},
\be
\label{t1}
\F^t=-T^4\, \Big( n_T^t\, \big( f_T(z) + k_T(z,\eta - |\zeta|)\big)+n^t_V\, \big(f_V(z) + k_V(z,\eta - |\zeta|)\big)\Big)\, ,
\ee
where $n_T^t$ and $n_V^t$ are defined in Eqs (\ref{ntu}) and (\ref{ntv}), respectively.

- Case (I.ii) : Since the twisted sector of the partition function is independent of $R_6$, the result reported in Eq. (\ref{F1}) remains valid for arbitrary $\zeta$,
\be
\label{t2}
\F^t=-T^4\, \Big( n_T^t\, f_T(z)+n^t_V\, f_V(z)\Big)\, .
\ee

- Case (II.i) : The result is as in the pure thermal case, Eq. (\ref{intpartut}), up to the R-charge dependent factor as in Eq. (\ref{F2}),
\be
\label{t3}
\F^t=-T^4\, n^t_T\left(1-{\b\eta\over 2}\right) \big( c_4+k(y)\big)\, ,
\ee
where
\be
\label{y}
y=z+\eta-\abs\zeta\abs
\ee
and the function $k$ defined in Eq. (\ref{kgtemp}) interpolates between the four dimensional regime where $\zeta$ is close to zero and the five dimensional one for large $\abs\zeta\abs$.

- Case (II.ii) : The twisted contribution being independent of $R_6$, the result reported in Eq. (\ref{F2}) remains valid for all $\zeta$,
\be
\label{t4}
\F^t=-T^4\, n^t_T\left(1-{\b\eta\over 2}\right)  c_4\, .
\ee

The free energy density $\F$ is the effective potential at finite temperature for the modulus $\zeta$, as can be seen from its equation of motion \cite{BEKP},
\be
\ddot \zeta+3H\dot \zeta+{\partial \F\over \partial \zeta} = 0\, .
\ee
For the backgrounds of type (I.i), this potential has been analyzed in detail in Ref. \cite{BEKP}  and presents five distinct phases. Our aim is to generalize these results to the other cases in Eq. (\ref{models}) and examine the similarities and discrepancies.

$\bullet$ {\bf Phase 1:} Let us define a neighborhood of $R_6=1$ by
\be
\abs\zeta\abs<{1\over 2R_0} \mbox{ and } {1\over 2R_4}\, .
\ee
In case (I.i), one finds $\zeta(t)$ can be stabilized at the origin $\zeta=0$, which is a local minimum of $\F$.
An RDS attractor exists if the model-dependent integer coefficients in $\F$ satisfy the inequality (I$a$) or (I$b$) in Eq. (\ref{Iabc}). However, note that the relevant spectrum at the self-dual point includes the vector multiplets of the $SU(2)$ enhanced symmetry point,
\be
\mbox{(I$a$.i)}~: ~{n^u_V+\hat n^u_V+n_V^t\over n^u_T+\hat n^u_T+n_T^t}\le-{1\over 15}~~\qquad  ~~\qquad\mbox{(I$b$.i)}~:~ -{1\over 15}<{n^u_V+\hat n^u_V+n_V^t\over n^u_T+\hat n^u_T+n_T^t}<0\, .
\ee
In the latter case, $z$ is stabilized. In the former one, $z$ freezes at an arbitrary large and negative value $z_0$ and the cosmology is better interpreted in five dimensions.\footnote{We have generalized the analysis of case (I$b$.i) that can be found in Sect. 3.1.1 of Ref.  \cite{BEKP} to case (I$a$.i). For any  constant $z\equiv z_0$ such that $e^{4z_0}\ll 1$, an RDS is found as long as a ``residual force" that scales as $\dis \left({R_0/ R_4}\right)^4$ is neglected. Taking into account this correction, Eq. (3.14)  in \cite{BEKP} gets an additional constant term (while $\tilde C=0$). It follows that $\o\varepsilon_{(z)}=  e^{3z_0} {c_4\over c_5}\left({n_V+\hat n_V\over n_T+\hat n_T}+{1\over 15}\right)<0$, which is  similar to Eq. (\ref{conv}), with the same conclusions.}  These conclusions remain identically true for the backgrounds of type (I.ii). This is due to the fact that the functions $k_T$ and $k_V$ that make the difference between Eqs (\ref{t1}) and (\ref{t2}) are exponentially suppressed when $\zeta$ is in phase 1 (given the fact that $R_0\gg 1$).

For the same reason, the models of type (II.i) and (II.ii) can be analyzed simultaneously since their only difference in Eqs (\ref{t3}) and (\ref{t4}) is the function $k$, which is also exponentially suppressed in phase 1. The analysis of \cite{BEKP} applies to these cases and one concludes that the above RDS attractor with both $\zeta$ and $z$ stabilized exists if  $n_V+\hat n_V<0$. This refers to the union of ranges of parameters named (II$a$), (II$b$) in Eq. (\ref{IIabc}) for the enhanced spectrum.

Expanding the free energy to quadratic order, we find that the mass of $\zeta$ goes as $m_{\zeta} \propto T \propto M$, where the exact coefficients may be computed in each case and is dependent on the critical value of $z$.  Thus we may stabilize all radii moduli with masses of the order the supersymmetry breaking scale.\footnote{Current work is in progress to show this generalizes to all moduli excluding the dilaton \cite{Estes:2011iw}.}  One also has that $\dot \zeta =\o \zeta H \ll H$ as the Universe expands so that energy stored in the oscillations of $\zeta$ dilutes faster than the thermal energy and the moduli do not dominate at late times.  It is only at the exit of the intermediate era, where the electro-weak symmetry breaking takes place, that $M$ and thus $m_\zeta$ reach their final values of order the electro-weak scale. We would like to stress that in models where one stabilizes the moduli with a constant mass from the outset, one always reaches an era where the Universe is dominated by moduli and no longer thermal like \cite{cosmomodprob}.  For the models considered in this paper, and in particular due to the way we break supersymmetry, the moduli may dilute their kinetic energy during the intermediate era so that the Universe is never moduli dominated even after $m_\zeta$ is fixed.

$\bullet$ {\bf Phase 2:} When $\zeta$ is in the range
\be
\label{phase2def}
{1\over 2R_0} \mbox{ and } {1\over 2R_4}<\zeta<\ln R_0 \mbox{ and }\ln R_4\, ,
\ee
the functions $k_T$, $k_V$, $k$ and $g_T$, $g_V$ are exponentially small and the thermal effective potential of $\zeta$ is flat. As in \cite{BEKP}, all conclusions found in the previous phase 1 apply to phase 2, up to two differences. First, $\zeta(t)$ freezes at an arbitrary value along the above flat direction. It is not stabilized but behaves at late times as a constant modulus. Second, since it sits away from the enhanced symmetry point, one has to replace $n^u_T+\hat n^u_T+n^t_T\to n^u_T+n^t_T$ and $n^u_V+\hat n^u_V+n_V^t\to n^u_V+n_V^t$.

$\bullet$ {\bf Phase 3:} When the inequality
\be
\zeta> \ln R_0 \mbox{ and } \ln R_4\, ,
\ee
is satisfied, the functions $k_T$, $k_V$, $k$ do contribute, while $g_T$ and $g_V$ are exponentially negligible. Actually, for $R_6\gg R_0$ and $R_4$, the free energy density depends on two variables, $z$ and $y=z+\eta-\abs\zeta\abs$ only. It is thus more natural to study the dynamics of $y$ instead of $\zeta$ \ie the relative motion of $R_6$ compared to $R_0$. To be specific, the untwisted contribution (\ref{fu}) takes the simplified form
\be
\label{Fun}
\F^u=- T^4\, \Big(e^{-y}\, \big(n_T^u\, f^{(5)}_T(z)+n_V^u\, f^{(5)}_V(z)\big)+e^{3y}\, (n_T^u+n_V^u){4\over 15}c_4\Big)\, ,
\ee
where
\be
\label{fTV5}
f^{(5)}_T(z)={\Gamma(3)\over \pi^{3}}\sum_{\tk_0,\tk_4}{e^{5z}\over \left[e^{2z}(2\tk_0+1)^2+(2\tk_4)^2\right]^{3}}\; , \qquad f^{(5)}_V(z)=e^{4z}\, f^{(5)}_T(-z)\, ,
\ee
while the twisted sectors yield,
\begin{eqnarray}
\label{1}&\!\!\!\!\!\!\!\!\!\!\!\!\!\!\!\!\!\!\!\!\mbox{case (I.i) }&: \quad  \F^t=- T^4\, \Big(e^{-y}\, \big(n_T^t\, f^{(5)}_T(z)+n^t_V\, f^{(5)}_V(z)\big)+e^{3y}\, (n_T^t+n_V^t){4\over 15}c_4+\cdots \Big)\\
\label{2}&\!\!\!\!\!\!\!\!\!\!\!\!\!\!\!\!\!\!\!\mbox{case (I.ii)  }&: \quad  \F^t=-T^4\, \Big( n_T^t\, f_T(z)+n^t_V\, f_V(z)\Big)\\
\label{3}&\!\!\!\!\!\!\!\!\!\!\!\!\!\!\!\!\!\!\mbox{case (II.i)  }&:{\dis  \quad  \F^t=-T^4\, n^t_T\left(1-{\b\eta\over 2}\right) }\Big( e^{-y}\, c_5+e^{3y}\, {8\over 15}c_4+\cdots \Big)\\
\label{4}&\!\!\!\!\!\!\!\!\!\!\!\! \!\!\!\!\!\mbox{case (II.ii)  }&:\dis  \quad \F^t=-T^4\, n^t_T\left(1-{\b\eta\over 2}\right)  c_4\, .
\end{eqnarray}
To reach these expressions, we have neglected terms of order $e^{-2\pi (R_0/R_6)^{-1}}$. In this regime, we find in all cases that the Universe is attracted back towards phase 2. That is, even though $R_6(t)$ is increasing, it turns out that $R_0(t)$ and $R_4(t)$ end by increasing faster and always catch $R_6(t)$ so that we enter into the regime of phase 2 as defined in (\ref{phase2def}).  This attraction is a result of the sub-dominant terms in the free energy density.\footnote{For the backgrounds where the direction 6 is a circle \ie for the cases (I.i) and (II.i), in the limit that one takes the initial value of $R_6$ to infinity or at least large enough so that one may neglect these sub-dominant terms, the evolution is attracted to an RDS in one more dimension.}  The interested reader can follow the detailed derivations and conclusions in each situation in Appendix F.

$\bullet$ {\bf Phase 4:} The dynamics in the range
\be
-\ln R_0 \mbox{ and }-\ln R_4<\zeta<{-{1\over 2R_0}}\mbox{ and }{-{1\over 2R_4}}\, ,
\ee
is identical to the one of phase 2, as follows from T-duality $\zeta\to -\zeta$.

$\bullet$ {\bf Phase 5:} Similarly, when $\zeta$ satisfies
\be
-\ln R_0 \mbox{ and } -\ln R_4<\zeta \, ,
\ee
which is T-dual to phase 3, the system is attracted back to phase 4.

\subsection{$\N$ = 1 $\to$ 0 models: Case (III)}

To extend the analysis of the internal space dynamics to models with $\N=1\to 0$, we reconsider the backgrounds (\ref{backg3}) of Sect. \ref{ZZ1}.  In this case, we remind the reader that the two $\Z_2$'s act on the directions 6,7,8,9 and 4,5,6,7, respectively.  The expression of the free energy density (\ref{FN=1}) is valid when the inequalities (\ref{hyp}) are valid. Our aim is to generalize it when one of the internal radii $R_I$ is arbitrary.  Two inequivalent situations must be considered, corresponding to the different ways the generators of the orbifold act. In the first case, the generators treat both  directions 4 and $I$  symmetrically, in which case $I=5$.  The second case corresponds to when the generators distinguish the  directions 4 and $I$, in which case we may take $I=6$ without loss of generality.\footnote{The $I=6,7$ and $I=8,9$ directions can be seen to be equivalent by noting that the product of the two generators of $\Z_2\times \Z_2$ acts on the directions 4,5,8,9.}  We label the cases as follows
\be
\label{models1}
\begin{array}{lll}
\mbox{case (III.i)}&: & S^1(R_0)\times T^{3}(R_{\rm box})\times\dis  {S^1(R_4)\times S^1(R_6) \times T^4\over \Z_2\times \Z_2}\\
\mbox{case (III.ii)} &:& S^1(R_0)\times T^{3}(R_{\rm box})\times\dis  {S^1(R_4)\times S^1(R_5) \times T^4\over \Z_2\times \Z_2}\, .\\
\end{array}
\ee

In the spirit of the previous section, we separate the free energy density in four ((un)twisted, (un)twisted) sectors, $\F=\F^{u,u}+\F^{t,u}+\F^{u,t}+\F^{t,t}$. Obviously, the $(u,u)$ contribution is simply half the result found for the $\N=2\to 0$ models, Eq. (\ref{fu}):
\be
\label{fu1}
\begin{array}{ll}
\F^{u,u}(z,\eta,\zeta) = - T^4 \, \Big( &\!\!\!\!n_T^{u,u} \, \big( f_T(z) + k_T(z,\eta - |\zeta|)\big) + n_V^{u,u}\,  \big(f_V(z) + k_V(z,\eta - |\zeta|)\big)\\
&\!\!\!\! + \, \hat n_T^{u,u} \, g_T(z,\eta,|\zeta|) + \hat n_V^{u,u} \, g_V(z,\eta,|\zeta|) \Big)
\end{array}\, ,
\ee
where
\be
\mbox{ case (III.i)}~ :~~ e^\zeta=R_6~~, \qquad~~~ \mbox{ case (III.ii)}  ~ :~~ e^\zeta=R_5\, .
\ee
The generic coefficients $n_T^{u,u}$ and $n_V^{u,u}$ are defined in Eq. (\ref{coef1}), while $\hat n_T^{u,u}=\hat n_T^{u,u}=2\cdot 2$ concern the additional contributions at the enhanced symmetry point $\zeta=0$. The other sectors are also easy to find:

- In case (III.i), each generator of the orbifold group acts non-trivially on the direction 6. Thus, $\F^{t,u}$ and $\F^{u,t}$ are independent of $\zeta$ and their expressions computed for $R_6$ close to one in Eq. (\ref{FN=1}) are valid for arbitrary $\zeta$. On the contrary, the product of the two $\Z_2$ generators acts non-trivially on the directions 4,5,8,9. The contribution $\F^{t,t}$ is then half the result found in the twisted sector of case (II.i), given in Eq. (\ref{t3}),
\be
\label{t6}
\F^{t,t}=-T^4\, n_T^{t,t}\, \left(1-{\b\eta\over 2}\right) \big( c_4+k(y)\big) \, .
\ee
The coefficient $n_T^{t,t}$ is given in Eq. (\ref{coef1}).

- In case (III.ii), it is only the second generator of the orbifold that acts non-trivially on the direction 5. Consequently,  $\F^{u,t}$ and $\F^{t,t}$ reported in Eq. (\ref{FN=1}) are valid for arbitrary $\zeta$. Finally, the remaining $(t,u)$ contribution is half the result found in the twisted sector of case (I.i), given in Eq. (\ref{t1}):
\be
\label{t5}
\F^{t,u}=-T^4\, \Big( n_T^{t,u}\, \big( f_T(z) + k_T(z,\eta - |\zeta|)\big)+n^{t,u}_V\, \big(f_V(z) + k_V(z,\eta - |\zeta|)\big)\Big)\, ,
\ee
where the coefficients $n_T^{t,u}$ and $n_V^{t,u}$ are given in Eq. (\ref{coef1}).

We observe that the formal expression of $\F$ in case (III.ii) is identical to the one encountered for the $\N=2\to 0$ models we denoted (II.ii) in the previous subsection. We conclude that the discussions and conclusions concerning the dynamics of $\zeta$ in phases 1 to 5 of the thermal effective potential are identical, \ie the final attraction is a four-dimensional RDS in phase 1, 2 or 4.

The dynamics in phases 1, 2 and 4 is common to cases (III.i) and (III.ii) \ie the evolution is attracted to an RDS in four dimensions. For the former background, we only need to discuss the behavior in phase 3. When $R_6\gg R_0$ and $R_4$, we have
\be
\begin{array}{l}
\F^{u,u}+\F^{t,t}=\dis -T^4 \left(e^{-y}\, \left[ n_T^{u,u}\, f^{(5)}_T(z)+n_V^{u,u}\, f^{(5)}_V(z)+n_T^{t,t}\left(1-{\b\eta\over 2}\right)c_5\right]+\O(e^{3y})+\cdots\right)\\
\F^{t,u}+\F^{u,t}=\dis -T^4\left( n_T^{t,u}\, f_T(z)+n^{t,u}_V\, f_V(z)+n_T^{u,t}\left(1-{\b\eta\over 2}\right)c_4\right):=-T^4\, \kappa(z)\, ,
\end{array}
\ee
where the dots denote terms of order $e^{-2\pi (R_0/R_6)^{-1}}$ we neglect. We observe that the free energy density in this regime is formally as in case (I$b$.ii) in the previous section, up to two differences. First, the appearance of the constant term $c_5$ makes the discussion simpler, \ie valid for any $n_V^{u,u}<0$. Second, the expression of $\kappa(z)$ contains an additional term $c_4$, which is positive. The conclusions are thus identical (see Appendix F): The evolution of the Universe is attracted towards phase 2, where we expect it to enter. Once there, it finally converges to an RDS in four dimensions.\footnote{Our analytic study of the attraction from phase 3 to phase 2 does not apply to the situations where $n_V^{u,u}+n_V^{t,u}<0$ with $n_V^{u,u}>0$, but we expect this fact to remain true.}


\section{Summary of results}
\label{Summ}

Working in the framework of perturbative string theory, we have studied aspects of the cosmology induced by finite temperature and spontaneous supersymmetry breaking by geometrical fluxes in one internal direction 4 of heterotic models. We have considered simple illustrative backgrounds in four dimensions, with initial $\N=1$ supersymmetry, as well as models with initial $\N=2$ supersymmetry.  The specific cases we considered are given in (\ref{backg1})--(\ref{backg3}) and the generic form of the free energy density we analyzed is of the universal form
\be
\label{d}
\F=-T^4\, \Big( n_T\, f_T(z)+\tn_T\, c_4+n_V\, f_V(z)\Big) \, ,
\ee
where  $e^z = M/T$ is the ratio of the supersymmetry breaking scale to the temperature. $\F$ is parameterized by three integers: The choice of initial background determines $n_T$ and $\tilde n_T$, while the details of the breaking of supersymmetry fixes $n_V$.  The models naturally divide into two types.  Models similar to type (I), with supersymmetry breaking flux wrapping a toroidal-like direction, will always have $\tn_T = 0$, while those similar to types (II) and (III), with supersymmetry breaking flux wrapping an orbifold-like direction, will always have $\tn_T > 0$.  For models of type (II) and (III), $\tn_T$ is related to the twisted sector that is independent of $R_4$.

Whenever $n_V / n_T > 0$, we find that the Universe is attracted to a phase of contraction where our quasi-static approximation breaks down.  The models of type (I) with $n_V/n_T \leq - 1/15$ induce a large hierarchy  $T/M\gg 1$.  The resulting evolution is well described by a purely thermal model in five dimensions.  For the remaining cases, we find that $z = \ln M/T$, is stabilized at the unique zero of the force in (\ref{r4p}), \ie $V_z (z_c) = 0$, and the evolution is attracted to an RDS in four dimensions.  The latter is radiation-like in the sense that one obtains the state equation, $\rho_{\rm tot} = 3 P_{\rm tot}$, only after one includes the contribution from the motion of $M(t)$.
We summarize the results as
\bea
\begin{array}{c|c|c|c}
   & \dis {n_V\over n_T} \leq -{ 1\over 15} & - \dis{1\over 15} < {n_V\over n_T} < 0 & \dis{n_V\over n_T} > 0 \\ \hline
  (I)         & RDS^5 & RDS^4 & {\rm contraction} \\
  (II), (III) & RDS^4 & RDS^4 & {\rm contraction}
\end{array}
\eea
In general, depending on  the low-energy particle spectrum and the initial space-time dimension (at the exit of the Hagedorn or inflation era), some internal directions with supersymmetry breaking flux may be dynamically forced to decompactify.
The analysis in \cite{cosmo2} confirms this possibility on models with geometrical flux in two internal directions.  It would be interesting to understand the full phase space for semi-realistic models that would prefer or at least admit four-dimensional RDS, and determine their corresponding spectrums.

In Sect. 5, we relaxed the frozen moduli restriction and allowed one of the radii $R_I$ $(I\neq 4)$  of the internal space to become dynamical.  In all cases, when the logarithm of this radius, $\zeta$, is in the range defined by
\be
\label{r}
\abs\zeta\abs <\ln R_0 \mbox{ and }\ln R_4\, ,
\ee
it is either dynamically stabilized at its self-dual point $\zeta = 0$, with a mass of order the supersymmetry breaking scale, or freezes at any value of the modulus-like phase defined by the range
\be
{1\over 2R_0} \mbox{ and } {1\over 2R_4}<|\zeta|<\ln R_0 \mbox{ and }\ln R_4\, .
\ee
In both cases we find that the energy stored in the modulus is always diluted faster than the thermal energy, thus avoiding the cosmological moduli problem.
When $\zeta > \ln R_0 \mbox{ and }\ln R_4$, we find that $R_0(t)$ and $R_4(t)$ always evolve so that $\zeta(t)$ enters back into the range (\ref{r}), where its motion halts.  The attraction to the modulus phase is stronger for orbifold-like directions as compared to toroidal-like directions.

\section{Conclusions}
\label{Conc}

The finite temperature stringy setup naturally suggests a separation of the cosmological evolution in at least four distinct phases, according to the value of the temperature. Namely:
\\$(i)$ The very early phase, or even the ``(Pre-)Big Bang phase'', where the underlying string degrees of freedom are excited, or even strongly coupled. Perhaps string dualities can be applied to understand this phase and resolve the naive classical Big Bang singularity \cite{stringgascosmo}.\\$(ii)$ The stringy Hagedorn phase, $T\simeq T_H$, where string oscillators and the thermal winding states must be properly taken into account. Both phases $(i)$ and $(ii)$ lead to a non-geometrical structure, e.g. the T-fold cosmologies studied recently in \cite{KTT}. In these high temperature, high curvature and high string coupling regimes, the topology and dimensionality of the space are not well-defined concepts.  Recent progress in understanding the Hagedorn phase has been made in \cite{GravFluxes,MassSusy,hybrid}.
\\ $(iii)$ The third phase is the focus of this paper and has features similar to that of a radiation-like Freedmann cosmology \cite{KP2,cosmo1,cosmo2,Bourliot:2009cx,Bourliot:2009na}. Here the Universe has cooled down to temperatures far below $T_H$ and the effects of string massive states are exponentially suppressed.\\$(iv)$ At lower temperatures, the effective field theory approach is valid. We are expecting new phenomena such as the electroweak phase transition, QCD confinement and structure formation to take place. We also expect  that in this phase, some dynamics becoming relevant at these low temperatures will stabilize the no-scale modulus $\Phi$ associated to the supersymmetry breaking scale \cite{Noscale}, realizing a cosmological, dynamical mechanism for the scale hierarchy, $M_W\ll M_{\rm Planck}$.

In $\N=1$ models, such a stabilization mechanism exists.  At late times, one additional scale which enters the problem is the infrared renormalisation group invariant transmutation scale, $Q$, which is induced at the quantum level by the radiative corrections of the soft supersymmetry breaking terms at low energies \cite{Noscale,NoscaleTSR}.  It is possible, when $T(t) \le  Q$, for the radiative corrections to generate the potential for the Higgs and induce the electroweak phase transition,  $SU(2)\times U(1)\to U(1)_{\rm em}$.  This starts to be the case at a time $t_W$, and for times afterwards, $t>t_W$, the supersymmetry breaking scale $M$ is stabilized at a value close to $Q$.  Whether the correct Higgs potential is generated depends on the initial data; however, this initial value problem is avoided thanks to the attractor mechanism towards the RDS in earlier cosmological times.  It would be interesting to realize this scenario explicitly in string theory, and to indeed identify models which produce the radiative symmetry breaking and also stabilize the supersymmetry breaking scale $M$.  It would be especially interesting to find semi-realistic models which produce supersymmetry breaking scales compatible with current observations.

In earlier cosmological times where $M(t),T(t)\gg Q$, the transmutation scale is irrelevant and does not modify our analysis.  The results of this paper are thus valid in the intermediate cosmological history, $t_E<t<t_W$.  These statements are correct if one assumes that there is no hidden sector gauge group $G$ that confines at an IR renormalization group invariant scale $\Lambda_G$ above $Q$. In string theory models with such a hidden sector, we expect the attractor mechanism of the intermediate era to be valid above and below $\Lambda_G$, with threshold effects around $\Lambda_G$ \cite{intradcor}.

We have shown the existence of models (still consistent with the quasi-static and perturbative hypothesis) which describe decompactifications of internal radii involved in the spontaneous breaking of supersymmetry.  In these special cases, the cosmology is attracted to radiation-like dominated solutions in higher dimension.  It would be interesting to generalize this mechanism and try to generate the spatial directions of our Universe.  We may consider a scenario where three of the spatial directions form a small three-torus, $T^3$.  In the cases where the $T^3$ is wrapped by supersymmetry breaking flux, it is possible for it to decompactify and generate the three-dimensional space.  One could try to realize this scenario explicitly within string theory and investigate the connections between the resulting low energy particle spectrum and the requirements to generate three-dimensional space.  One candidate for studying this mechanism, when one direction has already decompactified, is the recent MSDS models \cite{MassSusy}.

In scenarios where the Universe starts out very small, close to the Planck or string scale, and taking the results of this paper into account, we find during the radiation-like era that while the internal radii (not involved in the supersymmetry breaking) may initially expand, they are always attracted to their flat potential phase where their evolutions halt.  This provides a natural mechanism for keeping moduli at or near the string scale and eventually stabilize them at enhanced symmetry points.

We have chosen our setup so that the underlying two-dimensional conformal field theories are exactly known in string length.   This restricted the supersymmetry breaking to occur via geometrical fluxes.  It would be interesting to use string-string dualities and re-interpret our results as non-perturbative effects in a dual theory.  This would allow us to understand the role of temperature in more general cases where supersymmetry is broken by non-perturbative effects.  For example, in the type II version of our setup there is perturbatively no enhancement of symmetry at the self-dual point.  However, a dual type II description of  the heterotic gauge group enhancement can be considered in terms of singularities in the internal space. For instance, a type IIA D2-brane wrapped on a vanishing ${\mathbb{CP}}^1$ cycle of radius dual to $R_6$ can give rise to an $SU(2)$ gauge theory and admits a mirror description in type IIB \cite{het/II}. The equivalence between the brane-world and geometrical singularity pictures can be analyzed along the lines of \cite{KP}.


\section*{Acknowledgements}

We are grateful to F. Bourliot, L. Liu and N. Toumbas for useful discussions.
H.P. thanks the Ecole Normale Sup\'erieure for hospitality.\\
\noindent This work is partially supported by the ANR (CNRS-USAR) 05-BLAN-0079-02 and CEFIPRA/IFCPAR 4104-2 contracts. The work of J.E. and H.P is also supported by the contracts ERC AdG 226371, PICS  3747 and 4172. J.E. acknowledges financial support from the Groupement d'Int\'er\^et Scientifique P2I.


\section*{Appendix A: Black body and Stefan's law revisited}
\renewcommand{\theequation}{A.\arabic{equation}}
\renewcommand{\thesection}{A}
\setcounter{equation}{0}
\label{stef}

In this appendix, our aim is to present our approach in the simplest context. Namely, we consider at the classical level supersymmetric string models in $D$-dimensional flat space-time.  At finite temperature, the supersymmetry is spontaneously broken and a cosmological evolution is induced at the quantum level. Restricting our analysis to the dynamics that follows the Hagedorn era, we have $T\ll T_H$. For simplicity, we first suppose that all other scales in the model are much higher than $T$. It follows that the only states that can be thermalized are massless and we recover Stefan's law and an attraction of the Universe towards a radiation dominated era.  Then, we show the above hypothesis on the scales of the model is actually a consequence of the dynamics of the internal space moduli.

From a statistical physics point of view, the system is the Universe filled with a thermal gas of states. To be concrete, the space is treated as a large box with periodic boundary conditions along its $D-1$ dimensions, which is nothing but a torus $T^{D-1}$. To regularize IR divergences, the sums over the KK states along these directions are replaced by continuous integrals. The supersymmetric spectrum of states which is thermalized is the whole set of string modes of a given model. This guaranties that the loop corrections are also free of UV divergencies.  From a dynamical point of view, the gas exerts a force on the ``walls of the spatial box'' and, if the induced  perturbation of its radius is small, a quasi-static evolution takes place.

To compute the canonical ensemble free energy, we consider Euclidean backgrounds of the form
\be
\label{backT}
S^1(R_0)\times T^{D-1}(R_{\rm box})\times \M_{10-D}\, ,
\ee
where $R_{\rm box}$ is the radius of the circles of $T^{D-1}$ and $\M_{10-D}$ is the internal space that preserves at least one supersymmetry in $D$ dimensions. $R_0$ is the radius of the Euclidean time circle along which bosons and fermions have different boundary conditions, so that all supersymmetries are spontaneously broken by thermal effects. To be specific, we consider heterotic models with $\M_{10-D}=T^{10-D}$, whose 1-loop partition functions take the form
\be
\label{Zth}
\begin{array}{ll}
Z=\dis R_0R_{\rm box}^{D-1}&\!\!\!\!\dis \int_F {d\tau_1d\tau_2\over 2\tau_2^{1+D/2}}{1\over 2}\sum_{a,b}(-)^{a+b+ab}{\theta[^a_b]^4\over \eta^4}
\, {\Gamma_{(10-D,10-D)}\Gamma_{(0,16)}\over \eta^{8}\, \bar\eta^{24}}\\
&\dis \sum_{n_0,\tm_0}e^{-{\pi R_0^2\over \tau_2}\abs \tm_0+n_0\tau\abs^2}(-)^{a\tm_0+bn_0+\tm_0n_0} .
\end{array}
\ee
In this expression, the $\Gamma$'s stand for Narain lattices. Orbifold models with $\M_{10-D}=T^{6-D}\times \dis {T^4/ \Z_2}$ have instead,
\be
\label{orbif}
\begin{array}{ll}
Z= R_0R_{\rm box}^{D-1}&\dis \!\!\!\!\int_F{d\tau_1d\tau_2\over 2\tau_2^{1+D/2}}\, \dis{1\over 2} \sum_{H,G}\, {1\over 2}\sum_{a,b}(-)^{a+b+ab}{\theta[^a_b]^2\theta[^{a+H}_{b+G}]\theta[^{a-H}_{b-G}]\over \eta^4}\\
&\dis{\Gamma_{(6-D,6-D)}\over \eta^4\bar\eta^{20}}\, Z_{(4,4)}^{(0,0)}[^H_G]\, {1\over 2}\sum_{\b\gamma,\b\delta}\b\theta[^{\b\gamma}_{\b\delta}]^8\, {1\over 2}\sum_{\b\gamma',\b\delta'}\b\theta[^{\b\gamma'}_{\b\delta'}]^6\b\theta[^{\b\gamma'+H}_{\b\delta'+G}]\b\theta[^{\b\gamma'-H}_{\b\delta'-G}]\\
&\dis \sum_{n_0,\tm_0}e^{-{\pi R_0^2\over \tau_2}\abs \tm_0+n_0\tau\abs^2}(-)^{a\tm_0+bn_0+\tm_0n_0},
\end{array}
\ee
where we define
\be
\label{Tn/Z2}
Z_{(n,n)}^{(0,0)}[^H_G]=\left\{
\begin{array}{ll}
\dis {\Gamma_{(n,n)}\over \eta^4\b\eta^4}&\dis \mbox{for } [^H_G]\equiv [^0_0],\\
\dis 2^n\left({\eta\b\eta\over \theta[^{1-H}_{1-G}]\b\theta[^{1-H}_{1-G}]}\right)^{n/2}&\dis\mbox{for } [^H_G]\not\equiv[^0_0].
\end{array}
\right.
\ee
 Due to the phase $(-)^{bn_0}$, the GSO projection in the odd winding sector is reversed and tachyons occur when $R_0$ reaches the Hagedorn radius, close to 1. As said before, we are interested in the low temperature regime (compared to $T_H$) where
\be
 R_0\gg 1\, ,
 \ee
so that $Z$ is well defined. Moreover,  we suppose for the moment that the contributions to the masses from the internal lattice are ``heavy". For instance, the internal radii $R_I$ satisfy
\be
\label{scale}
{1\over R_0}\ll R_I\ll R_0\, .
\ee
Following the steps detailed in Ref. \cite{cosmo1} (or appendix B for more sophisticated models where supersymmetry is spontaneously broken even at zero temperature), the partition function (\ref{Zth}) becomes
\be
\label{Zstef}
Z={R_0R_{\rm box}^{D-1}\over R_0^D}\, n_T \, c_D\qquad \where \qquad c_D={\Gamma(D/2)\over \pi^{D/2}}\sum_{\tk_0}{1\over \abs 2\tk_0+1\abs^D}\, ,
\ee
where $n_T$ is the number of massless boson-fermion pairs in the supersymmetric parent model,  when the temperature is not switched on. These states give rise to KK towers of modes along the Euclidean time circle (the discrete sum on $\tilde k_0$ is obtained by Poisson resummation of the KK momenta along $S^1(R_0)$).
In this expression, we have neglected exponentially small terms of order $\O(e^{-2\pi R_0\,Mass})$, where $Mass$ stands for a mass contribution arising from oscillators and/or internal lattice zero modes. The integer number $n_T$ depends on the specific values taken by the internal moduli. In the toroidal case (\ref{Zth}), on has $n_T=8\, (504+N_{\rm enhan})$, where $N_{\rm enhan}$ stands for the additional contribution that arises at an enhanced symmetry point. In the orbifold models (\ref{orbif}), $n_T$ refers to untwisted and twisted modes, $n_T=4 \,(504+N_{\rm enhan}+512)$.

In the regime we study, the notion of space-time is well defined in field theory (the radius of curvature is large in all $D$ dimensions) and the dynamics at low energy can be described by an effective action $S$ in Lorentzian time. The first non-trivial contribution to the vacuum energy arises at genus one. Supposing the string coupling $e^\phi$ in $D$ dimensions is small enough for perturbation theory to be valid, we can write $S$ at 1-loop order,
\be
S=\int d^Dx \sqrt{-g_{\rm st}} \left[e^{-2\phi}\left({R_{\rm st}\over 2}+2(\partial \phi)^2+\cdots\right)+{Z\over \beta V_{\rm box}}\right],
\ee
for the string frame metric $g_{{\rm st}\mu\nu}$ and the dilaton. The dots stand for the other massless degrees of freedom, while the massive states are integrated out.\footnote{The 1-loop corrections to the kinetic terms can be absorbed by wave function renormalization. They would translate into corrections to the vacuum energy at second order only.} In this expression, we denote $\beta=2\pi R_0$ and $V_{\rm box}=(2\pi R_{\rm box})^{D-1}$.
For $D\ge 3$, the action can be rewritten in the Einstein frame as,
\be
S=\int d^Dx \sqrt{-g} \left[{R\over 2}-{1\over 2}(\partial \phi_\bot)^2+\cdots-\F \right]\, ,
\ee
where we have defined
\be
\label{Fstef}
\F=-T^D\,  n_T\,  c_D\; , \qquad T={1\over 2\pi R_0 \, e^{-{2\phi\over D-2}}}\; , \qquad \phi_\bot={2\over \sqrt{D-2}}\phi\, .
\ee
Supposing the back-reaction of the 1-loop source $\F$ on the classical space-time induces a quasi-static evolution, we look for homogeneous and isotropic extrema to $S$, with metric and dilaton ansatz,
\be
\label{ansatz}
\begin{array}{l}
ds^2=-N(t)^2dt^2+a(t)^2\left[(dx^1)^2+\cdots+ (dx^{D-1})^2\right]\; , \qquad \phi_\bot(t)\, , \\
\where \displaystyle\qquad N(t)\equiv 2\pi R_0 \, e^{-{2\phi\over D-2}}\equiv {1\over T(t)}\; ,\qquad a(t)\equiv 2\pi R_{\rm box} \, e^{-{2\phi\over D-2}}\, ,
\end{array}
\ee
and trivial background for the other massless fields.
Note that the laps function is by construction the inverse temperature, since it is obtained by analytic continuation from the Euclidean background.  The components of the stress-energy tensor ${T_{{\rm tot}\mu}}^\nu={\rm diag}{(-\rho_{\rm tot},P_{\rm tot},P_{\rm tot},P_{\rm tot})_\mu}^\nu$ satisfy
\be
\label{Prhotot}
P_{\rm tot}={T^2\over 2}\, \dot\phi_\bot^2-\F\; ,\quad \rho_{\rm tot}={T^2\over 2}\, \dot\phi_\bot^2+\F-T\, {\partial \F\over \partial T}\quad  \Longrightarrow\quad \rho_{\rm tot}=-P_{\rm tot}+T\, {\partial P_{\rm tot}\over \partial T}.
\ee
Separating the classical and 1-loop contributions, $\rho_{\rm tot}=\rho_\bot+\rho$,  $P_{\rm tot}=P_\bot+P$, the system $(\rho, P)$ satisfies Stefan's law, the state equation for radiation in $D$ dimensions,
\be
\rho=(D-1)P=(D-1)\, T^D\, n_T\, c_D\, .
\ee

Note that for more general expressions of $\F$, we  recover from the variational principle the usual quantum statistical results
\be
\label{Prho}
P=-e^{2D\phi\over D-2}\left({\partial F_{st}\over \partial V_{\rm box}}\right)_\beta\; ,\qquad \rho={e^{2D\phi\over D-2}\over V_{\rm box}}\left({\partial (\beta F_{st})\over \partial \beta}\right)_{V_{\rm box}},
\ee
where $F_{st}$ is the free energy associated to the thermal partition function $\dis {\cal Z}= {\rm Tr} \, e^{-\beta H}$,
\be
F_{st}=-{\ln {\cal Z}\over \beta}= -{Z\over \beta}.
\ee

Redefining a more conventional time variable corresponding to a constant laps function equal to 1, the Friedmann and $\phi_\bot$ equations of motion are,
\ba
\label{Fri}&&\dis {1\over 2}(D-1)(D-2)\, H^2={1\over 2}\dot\phi_\bot^2+\rho \, ,\qquad \left( H\equiv {\dot a\over a}\right),\\
&&\dis \ddot \phi_\bot+(D-1)H\dot\phi_\bot=0\qquad \Longrightarrow \qquad \dot\phi_\bot={c_{\bot}\over a^{D-1}}\, ,
\ea
where $c_\bot$ is a constant.
The conservation of the stress-energy  tensor  can be used to relate the temperature to the scale factor,
\be
\dot\rho+(D-1)H(\rho+P)=0\qquad \Longrightarrow\qquad aT=a_0T_0\, ,
\ee
where $a_0T_0$ is a constant.
The Friedmann equation (\ref{Fri}) can be used to show that $a(t)\to +\infty$ as $t\to+\infty$. This implies that for late times, the classical kinetic energy density $\dis \rho_\bot\propto {1\over a^{2(D-1)}}$ is negligible, compared to the thermal one $\dis \rho\propto {1\over a^D}$. Therefore, for arbitrary IBC at the exit time $t_E$ of the Hagedorn era, the cosmological evolution  is attracted to a radiation dominated era, since
\be
\label{rds1}
{1\over 2}(D-1)(D-2)\, H^2= {C_r\over a^D} \qquad \where\qquad C_r=(D-1)(a_0T_0)^Dn_Tc_D\, ,
\ee
and the fields satisfy
\be
\label{rds2}
a(t)=t^{2/D}\times \left({2C_r\over (D-1)(D-2)}\right)^{1/D}={1\over T(t)}\times (a_0T_0)\; , \qquad \phi_\bot=cst.
\ee

\smallskip

\noindent {\large \em Stabilization of the internal space}

\noindent
At this stage, we have supposed the space $\M_{10-D}$ in (\ref{backT}) is static and all scales determined by the internal moduli are far above the temperature. We want here to examine if these conditions are reached dynamically. For concreteness, we consider the orbifold model (\ref{orbif}) with $D=4$ and internal space $\M_6=T^2\times \dis {T^4/\Z_2}$.
Our aim is to study the evolution of the moduli of this space where, for simplicity, we allow only one radius of either $T^2$ or the orbifold factor to be arbitrary and dynamical. In both cases, we denote this radius as $R_6$, while keeping the other moduli frozen and such that the associated scales they define are much larger than $T$. Precisely, the internal space is taken to be either
\be
\M_6=S^1(R_6)\times S^1\times {T^4\over \Z_2}\qquad~~ \mbox{or}\qquad ~~T^2\times {S^1(R_6)\times T^3\over \Z_2}\, ,
\ee
and for instance the $R_I$'s in all internal directions but 6 satisfy the inequality (\ref{scale}).
For convenience, we introduce the notations
\be
e^{\tau}=R_0\; , \qquad e^{\zeta}=R_6\, .
\ee

In the case $R_6$ is a radius of $T^2$, we use the general analysis of appendix A in Ref. \cite{BEKP} to write the free energy density for arbitrary $\zeta$ as,
\be
\label{intpartut}
\F =  - T^4\,  \Big( (n_T^u+n_T^t)\, \big(c_4+k(\tau-\abs\zeta\abs)\big)+\hat n_T^u \, g(\tau,\abs\zeta\abs)\Big)\, .
\ee
In this expression, $n^u_T=4\left[504+N_{\rm enhan}\right]$ and $n^t_T=4\cdot 512$ are the numbers of massless boson-fermion pairs in the untwisted and twisted sectors (at zero temperature), when $R_6$ takes a generic value.\footnote{The integer $N_{\rm enhan}$ refers to the additional massless contributions that may occur at  specific points of the space of internal moduli other than $R_6$.} However, an $SU(2)$ enhancement of the gauge symmetry arises at $R_6=1$. The additional massless states that arise at this point are taken into account by the term proportional to $\hat n^u_T=4\cdot 2$. In Eq. (\ref{intpartut}), the functions we have introduced are
\be
\label{kgtemp}
\begin{array}{rl}
&k(\tau-\abs\zeta\abs) =\dis  e^{2(\tau-\abs\zeta\abs)}\,  {\sum_{m_6}}' \sum_{k_0} {2\, m_6^2  \over (2 k_0 + 1)^2} \,  K_2 \Big( 2 \pi |(2 k_0 + 1) m_6| e^{\tau-\abs\zeta\abs}\Big) \, ,\\
&g(\tau,\abs\zeta\abs) = \dis e^{2\tau}\, \left(e^{\abs\zeta\abs}-e^{-\abs\zeta\abs}\right)^2\,  \sum_{k_0} {2\over (2 k_0+1)^2} \,  K_2 \Big( 2 \pi |2 k_0 + 1| e^\tau \left(e^{\abs\zeta\abs}-e^{-\abs\zeta\abs}\right)\!\Big)\, ,
\end{array}
\ee
where $K_2$ are modified Bessel functions of the second kind.
They depend on $\abs\zeta\abs$ only, as a consequence of the T-duality $R_6\to \dis {1/ R_6}$ of the model.

When $R_6$ is one of the radii of $\dis {T^4/ \Z_2}$, the untwisted contribution is as in Eq. (\ref{intpartut}). Since the twisted sector is independent of $R_6$, the contribution to the free energy density can be computed for $R_6$ satisfying Eq. (\ref{scale}). The net result is,
\be
\label{bis}
\F =  - T^4\,  \Big( n_T^u\, \big(c_4+k(\tau-\abs\zeta\abs)\big)+\hat n_T^u \, g(\tau,\abs\zeta\abs)+n_T^t\, c_4\Big)\, .
\ee

To discuss the dynamics of $R_6(t)$ in both cases, we may consider a general free energy density of the form
\be
\label{intparttot}
\F =  - T^4\,  \Big( n_T\, \big(c_4+k(\tau-\abs\zeta\abs)\big)+\hat n_T \, g(\tau,\abs\zeta\abs)+\tn_T\, c_4\Big)\, ,
\ee
where the main difference is that $\tilde n_T = 0$ if $R_6$ is a radius of $T^2$ and $\tilde n_T > 0$ if it is a radius of the orbifold factor.
Our interest in $\F$ is justified by the fact that it plays the role of an effective potential at finite temperature for $R_6$, as follows from the equation of motion for $\zeta$ (see appendix B of \cite{BEKP}),
\be
\label{inteom}
\ddot \zeta + 3 H \dot \zeta + {\partial \F \over \partial \zeta} = 0\, .
\ee
As shown on Fig. \ref{thermal_potential}, this potential presents five phases. They can be analyzed along the lines of \cite{BEKP,Estes:2011iw}:
\begin{figure}[h!]
\begin{center}
\vspace{.3cm}
\includegraphics[height=5.5cm]{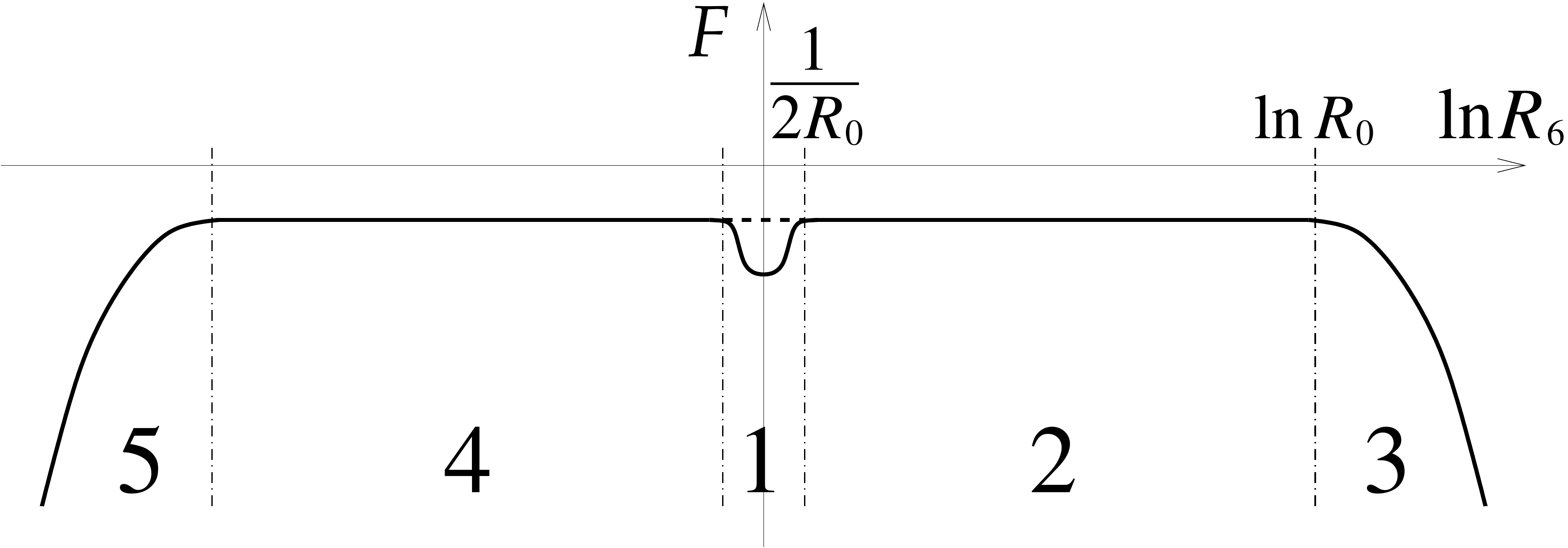}
\caption{\footnotesize \em  Thermal effective potential for the scalar $\zeta=\ln R_6$, at $\tau=\ln R_0$ and temperature $T$ constant. It is nothing but the free energy density $\F$.}
\vspace{-.4cm}
\label{thermal_potential}
\end{center}
\end{figure}

$\bullet$ {\bf Phase 1:} In the neighborhood of $\zeta=0$ defined by $\abs\zeta\abs<\dis {1\over 2R_0}$, {we have $k=0$ in Eq. (\ref{intparttot}), up to exponentially suppressed terms, thanks to $R_0\gg 1$. We also have $g(\tau,0)=c_4$ and $\F$ admits a local minimum for $\zeta$, since
\be
\label{quadra}
\F= - T^4\, \Big ( (n_T + \hat n_T+\tilde n_T) \, c_4 - \hat n_T \, \pi^2\, e^{2\tau}\, \zeta^2 + {\cal O}(\zeta^4)\Big)\, .
\ee
It follows that Eq. (\ref{inteom}) takes the form of a damped harmonic oscillator, with time dependent coefficients, and qualitatively one may expect that $\zeta$ is attracted to the solution $\zeta = 0$ for an expanding Universe.  To make the argument more precise, we may consider a general perturbation around the point $\zeta = 0$ along the lines of \cite{Estes:2011iw}.  Under such a perturbation, we find the evolution is attracted back to the point $\zeta = 0$, while the cosmological evolution is ``radiation-like".   What is meant by this is that the scale factor evolution is that of a radiation dominated Universe but the total energy density contains contributions from both thermal radiation and from the moduli. The fluctuation in $\zeta$ sources the dilaton, resulting in a logarithmic behavior towards weak coupling and so $\phi_\bot$ does not asymptote to a constant finite value.  Note that in higher dimensions, the Universe is radiation dominated and that $\phi_\bot$ asymptotes to a constant finite value.

$\bullet$
{\bf Phase 2:} When $\dis {1 \over 2 R_0}<\zeta < \tau$, both functions $k$ and $g$ in Eq. (\ref{intparttot}) are exponentially suppressed and we have
\be
\F= - T^4\, (n_T +\tilde n_T) \, c_4\, .
\ee
Thus, any constant $\zeta(t)$ in the above range solves Eq. (\ref{inteom}) and one finds that the Universe can again be attracted  to a radiation dominated era. However, $\zeta$ is not stabilized as before. It behaves as a modulus frozen by the friction term arising from the expansion of the Universe in Eq.  (\ref{inteom}). Since $R_6(t)$ is freezing while $R_0(t)$ increases, the inequalities (\ref{scale}) are better and better satisfied, even for $I=6$.

$\bullet$
{\bf Phase 3:} When $\zeta > \tau$, the function $g$ in Eq. (\ref{intparttot}) is exponentially suppressed and it is more convenient to study the dynamics of the field
\be
y=\tau-\abs\zeta\abs
\ee
instead of $\zeta$. Using the identity
\be
c_4+k(\tau-\abs\zeta\abs)=h(y):={\Gamma\left({5/ 2}\right)\over \pi^{5/2}}\sum_{\tk_0,\tm_6}{e^{4y}\over \left[e^{2y}(2\tk_0+1)^2+\tm_6^2\right]^{5/ 2}}\, ,
\ee
which is obtained by  Poisson resummation on $m_6$ in the definition of $k$, we have for $y<0$
\be
\F=-T^4\, \Big( n_T\, h(y)+\tn_T \, c_4\Big):= -T^4\, p(y)\, .
\ee
The coupled equations of motion for the temperature, scale factor, dilaton and $y$ are identical to the one considered in the problem solved in Sect. \ref{radera} and given in Eqs (\ref{Ta})--(\ref{syseq}) (together with the definitions (\ref{pr}) and (\ref{r4p})) under the replacement $z\to y$. In particular, the field $y$ has a potential whose derivative
\be
V_y = - n_T \, {5\, \Gamma\left({5\over 2}\right)\over \pi^{5\over 2}}  \sum_{k_0,\tilde m_6} {\tilde m_6^2 \, e^{4 y} \over [e^{2y}(2 \tk_0+1)^2 + \tilde m_6^2]^{7\over 2}}
- \tilde n_T \, c_4
\ee
is negative for any $y<0$. We find that there is always a force pushing $y$ towards larger values, meaning that $R_0(t)$ catches $R_6(t)$ when the Universe expands. In other words, the system is attracted to phase 2, where $y>0$ and $\zeta(t)$ starts freezing.  We justify this in two steps. First, note that in the limit $e^y\ll 1$, one has $V_y=-n_T \, e^{3y}\,  32c_4/15-\tn_T c_4$. This shows that the force towards the phase $y>0$ is essentially constant  when $\tn_T>0$ and much milder when $\tn_T=0$.  More generally, for arbitrary IBC at $t_E$ such that $y<0$ (but not necessary $e^y \ll 1$), we have completed our analytic study by a numerical simulation that confirms that $y$ is always driven to phase 2.
However, when $\tn_T=0$ and one chooses the initial value of $R_6$ to be infinite or sufficiently large, we can neglect the effect of this force.  Then, the dynamics is better understood in higher dimension, where $e^y\ll 1$  behaves as a modulus frozen by the friction term arising from the expansion of a radiation dominated five-dimensional Universe. $R_0(t)$ and $R_6(t)$ are then running away, proportionally to $R_{\rm box}(t)$. The proof of these statements is identical to the one given for the problem treated in Sect. \ref{T>M}, Eqs (\ref{z<-1})--(\ref{att}), under the replacement $R_4\to R_6$, $z\to y$, $n_V\to n_T$.

$\bullet$
{\bf Phase 4:} The dynamics in the regime $\dis -\tau <\zeta < -{1 \over 2 R_0}$ is common to phase 2, as follows from T-duality $\zeta\to -\zeta$.

$\bullet$
{\bf Phase 5:} The dynamics for $\dis  \zeta < -\tau$ is T-dual to that of phase 3 and the system is attracted back to phase 4.


\section*{Appendix B}
\renewcommand{\theequation}{B.\arabic{equation}}
\renewcommand{\thesection}{B.}
\setcounter{equation}{0}

Here, we give a summary of the derivation of the free energy density in models whose Euclidean heterotic background is given in (\ref{backg1}) and considered in \cite{cosmo1, BEKP}. The partition function is
\ba
\nonumber Z=&&\!\!\!\!\!\!\!\!\!\dis R_0R_{\rm box}^3R_4 \int_F{d\tau_1d\tau_2\over 2\tau_2^{7/2}}\dis{1\over 2}\sum_{H,G}{1\over 2}\sum_{a,b}(-)^{a+b+ab}{\theta[^a_b]^2\theta[^{a+H}_{b+G}]\theta[^{a-H}_{b-G}]\over \eta^4}\\
&&\dis \!\!\!\!\!\!\! \! \!{1\over 2}\sum_{\b\gamma,\b\delta}\b\theta[^{\b\gamma}_{\b\delta}]^8\, {1\over 2}\sum_{\b\gamma',\b\delta'}\b\theta[^{\b\gamma'}_{\b\delta'}]^6\b\theta[^{\b\gamma'+H}_{\b\delta'+G}]\b\theta[^{\b\gamma'-H}_{\b\delta'-G}]\,  {\Gamma_{(1,1)}\over \eta^4\bar\eta^{20}}\, Z_{(4,4)}^{(0,0)}[^H_G]\\
\nonumber &&\dis \!\!\!\!\!\!\! \! \!\sum_{n_0,\tm_0}e^{-{\pi R_0^2\over \tau_2}\abs \tm_0+n_0\tau\abs^2}(-)^{a\tm_0+bn_0+\tm_0n_0}\sum_{n_4\tm_4}e^{-{\pi R_4^2\over \tau_2}\abs \tm_4+n_4\tau\abs^2}(-)^{(a+\b Q_4)\tm_4+(b+\b L_4)n_4+\b\epsilon_4 \tm_4n_4},
\ea
where $\b Q_4, \b L_4,\b \epsilon_4$ are defined in Eq. (\ref{charges}) and $Z_{n,n}^{(0,0)}$ is given in Eq. (\ref{Tn/Z2}).
Redefining $n_i=2l_i+h_i$, $\tm_i=2\tk_i+\tg_i$, ($h_i, \tg_i=0, 1$ for $ i=0,4$) and $a=\hat a+h_0+h_4$, $b=\hat b+\tg_0+\tg_4$, the use of Jacobi identity yields
\ba
\nonumber Z=-\dis R_0R_{\rm box}^3R_4 \int_F{d\tau_1d\tau_2\over 2\tau_2^{7/2}}&&\!\!\!\!\!\!\!\!\!\!\dis{1\over 2}\sum_{H,G}{1\over 2}\sum_{\b\gamma,\b\delta}\b\theta[^{\b\gamma}_{\b\delta}]^8\, {1\over 2}\sum_{\b\gamma',\b\delta'}\b\theta[^{\b\gamma'}_{\b\delta'}]^6\b\theta[^{\b\gamma'+H}_{\b\delta'+G}]\b\theta[^{\b\gamma'-H}_{\b\delta'-G}]\,  {\Gamma_{(1,1)}\over \eta^4\bar\eta^{20}}\, Z_{(4,4)}^{(0,0)}[^H_G]\\
&&\dis\!\!\!\!\!\!\!\!\!\!\!\!\!\!\!\!\!\!\!\!\!\!\!\!\!\!\!\!\!\!\!\!\ \!\!\!\!\!\!\!\! \! \!\sum_{h_0,\tg_0,h_4,\tg_4}{\theta[^{1+ h_0+h_4}_{1+\tg_0+\tg_4}]^2\theta[^{1-h_0-h_4-H}_{1-\tg_0-\tg_4-G}]\theta[^{1-h_0-h_4+H}_{1-\tg_0-\tg_4+G}]\over \eta^4}\\
\nonumber &&\!\!\!\!\!\!\!\!\!\!\!\!\!\!\!\!\!\!\!\!\!\!\!\!\!\!\!\!\!\!\!\!\!\!\!\!\!\!\!\sum_{l_0,\tk_0,l_4,\tk_4}e^{-{\pi \over \tau_2}\sum_i R_i^2\abs (2\tk_i+\tg_i)+(2l_i+h_i)\tau\abs^2}(-)^{\sum_i (h_i+\tg_i)+\sum_{i,j} h_i\tg_j+h_0\tg_0+\b\epsilon_4h_4\tg_4+\b Q_4\tg_4+\b L_4 h_4}.
\ea
In the intermediate cosmological era, we have $R_i\gg 1$ ($i=0,4$) and there is no  Hagedorn-like singularity. Thus, the contributions with non-trivial winding numbers in the directions 0 and 4 are exponentially suppressed \ie we can keep the sectors with $h_i=l_i=0$ only. Among them, the sub-sectors with $\tg_0+\tg_4=0$ are supersymmetric and do not contribute (as seen from the presence of $\theta[^1_1]$ functions). Due to factors of order $e^{-\pi{R_i^2/ \tau_2}}$ in the integrand, the substantial contributions in the $\tau_2$-integral arise for $\tau_2\to +\infty$, up to exponentially suppressed terms in $R_i$.  This means that  the integration over the fundamental domain can be replaced by the sum over the entire upper half strip. Altogether, one obtains
\ba
\nonumber Z=\dis R_0R_{\rm box}^3R_4 \int_{-1/2}^{1/2}d\tau_1\int_0^{+\infty} {d\tau_2\over 2\tau_2^{7/2}}&&\!\!\!\!\!\!\!\!\!\!\dis{1\over 2}\sum_{H,G}{1\over 2}\sum_{\b\gamma,\b\delta}\b\theta[^{\b\gamma}_{\b\delta}]^8\, {1\over 2}\sum_{\b\gamma',\b\delta'}\b\theta[^{\b\gamma'}_{\b\delta'}]^6\b\theta[^{\b\gamma'+H}_{\b\delta'+G}]\b\theta[^{\b\gamma'-H}_{\b\delta'-G}]\,  {\Gamma_{(1,1)}\over \eta^4\bar\eta^{20}}\, Z_{(4,4)}^{(0,0)}[^H_G]\\
&&\dis\!\!\!\!\!\!\!\!\!\!\!\!\!\!\!\!\!\!\!\!\!\!\!\!\!\!\!\!\!\!\!\!\!\!\!\!\!\!\!\!\!\!\!\!\!\!\!\!\!\!\!\!\!\!\ \!\!\!\!\!\!\!\! \! \!{\theta[^1_0]^2\theta[^{1-H}_{\phantom{1}-G}]\theta[^{1+H}_{\phantom{1+}G}]\over \eta^4}\sum_{\tg_0+\tg_4=1}(-)^{\b Q_4\tg_4}\sum_{\tk_0,\tk_4}e^{-{\pi \over \tau_2}[R_0^2 (2\tk_0+\tg_0)^2+R_4^2(2\tk_4+\tg_4)^2]}.
\ea
The last sum over $\tk_0,\tk_4$ arises (by Poisson resummation)  from the KK towers of states associated to the directions 0 and 4.
Expanding the rest of the integrand in series of $q^A \b q^B$, the $\tau_1$-integral implements the level matching condition. The change of variable  $\tau_2=x\pi [R_0^2 (2\tk_0+\tg_0)^2+R_4^2(2\tk_4+\tg_4)^2]$ shows that the monomials $e^{-2\pi\tau_2(A+B)}$ with non vanishing $A+B$ are exponentially small, compared to the massless contributions with $A+B=0$. Note that in the untwisted sector $H= 0$, there are two kinds of such states. Some are present at any point in the moduli space of $S^1\times T^4/\Z_2$, while additional ones arise if we sit at some enhanced symmetry point, where states with non-trivial winding modes in the directions $I=5,\cdots,9$ are exceptionally massless.
The final result for the free energy density Eq. (\ref{FZ}) measured in Einstein frame is given in Eq. (\ref{F1}).


\section*{Appendix C}
\renewcommand{\theequation}{C.\arabic{equation}}
\renewcommand{\thesection}{C.}
\setcounter{equation}{0}

Let us derive the orbifold block $Z_{(1,1)}^{(a+\b Q_4,b+\b L_4)}[^H_G]$ given in Eq. (\ref{Z11}). The form of  $Z_{(1,1)}^{(a+\b Q_4,b+\b L_4)}[^0_0]$ can be understood from its analogue in a pure KK field theory. In this context, any field is multiplied by a phase $(-)^{(a+\b Q_4)\tm_4}$ when $x^4\to x^4+2\pi R_4\tm_4$. In string theory, each KK mode is the lowest of a tower of winding states and the generalized phase is found by imposing modular invariance,
\ba
Z_{(1,1)}^{(a+\b Q_4,b+\b L_4)}[^0_0]\!\!\!&=&\!\!\!{1\over\eta\b\eta} {R_4\over \sqrt{\tau_2}}\sum_{n_4\tm_4}e^{-{\pi R_4^2\over \tau_2}\abs \tm_4+n_4\tau\abs^2}(-)^{(a+\b Q_4)\tm_4+(b+\b L_4)n_4+\b\epsilon_4 \tm_4n_4}\\
\label{Z2a}&=&\!\!\!{1\over\eta\b\eta}\sum_{n_4,m_4}(-)^{n_4(b+\b L_4)}\, q^{{1\over 4}p_L^2}q^{{1\over 4}p_R^2}\; ,\quad p_{L,R}={m_4-{a+\b Q_4+\b \epsilon_4 n_4\over 2}\over R_4}\mp n_4R_4\\
\label{Z3}&=&\!\!\! {\rm Tr}\left[ (-)^{n_4(b+\b L_4)}\, q^{{1\over 4}p_{L\rm tot}^2}q^{{1\over 4}p_{R\rm tot}^2}\right].
\ea
The second line (\ref{Z2a}) is obtained by Poisson resummation and involves shifted momenta, while in (\ref{Z3}), $p_{L\rm tot}$ and $p_{R\rm tot}$ include the contributions of the bosonic oscillators in the direction 4.

The contribution of the sector $[^H_G]\equiv [^0_1]$ is found by inserting the $\Z_2$-generator $g$ into the trace. Since $g$ acts on the zero modes $p_{L,R}$ (and the oscillators) as a minus sign, a state contributing to the trace must have zero winding and shifted momentum \ie $n_4=0$, $m_4-{a+\b Q_4+\b \epsilon_4 n_4\over 2}=0$.  We thus have,
\ba
Z_{(1,1)}^{(a+\b Q_4,b+\b L_4)}[^0_1]\!\!\!&=&\!\!\!\delta_{a+\b Q_4,0\, \rm mod\, 2}\sum_{\rm oscillators} \left[ (-)^{N_4+\b N_4}\, q^{{1\over 4}p_{L\rm tot}^2}q^{{1\over 4}p_{R\rm tot}^2}\right]\, ,\\
&=&\!\!\! \delta_{a+\b Q_4,0\, \rm mod\, 2}\, 2\sqrt{\eta\b\eta\over \theta[^{1}_{0}]\b\theta[^{1}_{0}]}\, ,
\ea
where $N_4, \b N_4$ are the number of left and right bosonic oscillators in the direction 4.

Finally, the blocks associated to the twisted sectors $[^H_G]\equiv [^{\, 1}_G]$ are found by modular invariance, as summarized in Eq. (\ref{Z11}). Note that the full string partition function can be found using a different reasoning. One can consider an $S^1(R_4)$ (with trivial boundary conditions) on which one acts with two $\Z_2$-actions, the former being free and the latter not freely acting. Although equal, the final result appears in a different form \cite{Gregori:1997hi}.


\section*{Appendix D}
\renewcommand{\theequation}{D.\arabic{equation}}
\renewcommand{\thesection}{D.}
\setcounter{equation}{0}

Our starting point is the same string theory backgrounds (I), (II) and (III) and partition functions that gave rise to the free energy in Eq. (\ref{F12}). However, we reconsider the effective action (\ref{S}) in terms of fields and a free energy density defined from a five dimensional point of view (we denote them with ``primes" to avoid any confusion with their four dimensional counterparts). Our goal is to find the equations of motion from this perspective. Specifically, we have
\be
\label{S5}
S=\int d^5x \sqrt{-g'} \left[{R'\over 2}-{1\over 2}(\partial \phi'_\bot)^2+\cdots-\F'\right],
\ee
where $\phi_\bot'$ is the ``normalized'' dilaton $\phi'$ and $\F'$ is the free energy divided by the volume in five dimensions, measured in Einstein frame,
\be
\phi_\bot':={2\over \sqrt{3}}\, \phi'\; , \qquad \F'=-{Z\over (e^{-{2\phi'\over 3}}2\pi R_0) (e^{-{8\phi'\over 3}}V_{\rm box}')}\; , \qquad V_{\rm box}'= (2\pi R_4) \, V_{\rm box}\, .
\ee
$\F'$ takes a form analogous to Eq. (\ref{def4}),
\be
\F'=-T^{\prime 5}\, p'(z)\qquad \where\qquad p'(z)=e^z\, p(z)\; , \quad  e^z= {R_0\over R_4}\; , \quad T'={1\over 2\pi R_0\, e^{-{2\phi'\over 3}}}.
\ee
The homogeneous but anisotropic metric ansatz in Einstein frame is now given by
\be
\label{ansatz5}
\begin{array}{l}
ds^{\prime 2}=-N'(t)^2dt^2+a'(t)^2\left[(dx^1)^2+ (dx^{2})^2+ (dx^{3})^2\right]+b(t)^2(dx^4)^2\; , \qquad \phi'_\bot(t)\, , \\ ~ \\
\where \displaystyle\quad N'(t)\equiv 2\pi R_0 \, e^{-{2\phi'\over 3}}\equiv {1\over T'(t)}\; ,\quad a'(t)\equiv 2\pi R_{\rm box} \, e^{-{2\phi'\over 3}}\; ,\quad b(t)\equiv 2\pi R_4 \, e^{-{2\phi'\over 3}}\, .
\end{array}
\ee
Varying $S$ with respect to this metric, the 1-loop contribution to the stress-energy tensor defines the thermal energy density and pressures, $T'_\mu {}^\nu = {{\rm diag}(\rho',P',P',P',P_4')_\mu}^\nu$, where
\ba
\nonumber&& \dis P'=T^{\prime 5}\, p'(z)\; , \quad P'_4=P'+b\, {\partial P'\over \partial b}=T^{\prime 5}\, (p'-p'_z)\; , \quad \rho'=-P'+T'\, {\partial P'\over \partial T'}=T^{\prime 5} \, r'(z)\, ,\\
\label{source}&&\dis \with \qquad e^z={1\over b\, T'}\; , \qquad  r'(z)= e^z \, r(z)=4p'-p'_z\, .
\ea
There are three independent Einstein's equations (for $\mu=\nu=0, 1, 4$)  and one equation for the scalar $\phi_\bot'$. For convenience, we consider linear combinations of them and write the conservation of the stress-energy tensor and  Friedmann's equation (we denote $H'=\dot a'/a$, $K=\dot b/b$),
\ba
\label{e2}&&\dot \rho'+(3H'+K)(\rho'+P')+Kb\, {\partial P'\over \partial b}=0\, ,\\
\label{e1}&&3(H^{\prime 2}+H'K)={1\over 2}\dot\phi_\bot^{\prime 2}+\rho'\, .
\ea
Defining $\dis e^\xi={b\over a'}$, we have $K\equiv H'+\dot\xi$  and we choose the last two equations to be
\ba
\label{e3}&&\ddot\xi+(3H'+K)\dot\xi=b\, {\partial P'\over \partial b}\, ,\\
\label{e4}&&\ddot\phi_\bot'+(3H'+K)\dot\phi'_\bot=0\, .
\ea
Proceeding as in \cite{Antoniadis:1986ke, Bourliot:2009na, BEKP}, we introduce derivatives $\dis \o y\equiv {dy\over d\ln a'}$ and use Eqs (\ref{source}) to recast Eqs (\ref{e2})--(\ref{e4}) in the form,
\ba
\label{e2'}&& \dis \left[r_z'(z)-5r'(z)\right]\o z+\left[p'(z)-4r'(z)-p_z'(z)\right]\o \xi=0,\\
\label{e1'}&&H^{\prime 2}=T^{\prime 5}\, {r'(z)\over 6+3\, \o\xi-{1\over 2}\, \o\phi{}_\bot^{\prime 2}},\\
\label{e3'}&&{r'(z)\over 6+3\, \o\xi-{1\over 2}\, \o\phi{}_\bot^{\prime 2}} \, \oo\xi+p'(z)\, \o\xi+p'_z(z)=0,\\
\label{e4'}&&{r'(z)\over 6+3\, \o\xi-{1\over 2}\, \o\phi{}_\bot^{\prime 2}}\,  \oo\phi{}_\bot'+p'(z)\, \o\phi{}_\bot'=0.
\ea


\section*{Appendix E}
\renewcommand{\theequation}{E.\arabic{equation}}
\renewcommand{\thesection}{E.}
\setcounter{equation}{0}

We consider here the equations of motion (\ref{e2'})--(\ref{e4'}), where the terms linear and quartic in $e^z=\dis {R_0/ R_4}$ are small compared to 1 and neglected in the partition function or free energy density (\ref{z<-1}). In this regime, our aim is to show that for arbitrary IBC, the two scalar fields $\xi(t)$ and $\phi'_\bot(t)$ converge to constants for late times.

\noindent {\small $\bullet$} For $\o\xi\not\equiv 0$, we define a new function $\chi$ such that $\o\phi{}'_\bot\equiv \chi\o\xi$. Using (\ref{e3'}) and the state equation (\ref{st5}), one finds immediately that $\chi$ is a constant and we need to solve the single equation
\be
\label{eq1}
4\, \oo\xi+\o\xi\left( 6+3\, \o\xi-{\chi^2\over 2}\, \o \xi{}^2\right)=0.
\ee

- When $\chi\neq 0$,  the solution to (\ref{eq1}) is,
\be
\label{sol}
\begin{array}{l}
\abs\o\xi\abs^{1\over r_+r_-}+(\o\xi-r_-)^{1\over r_-(r_--r_+)}+(r_+-\o\xi)^{1\over r_+(r_+-r_-)}=\left(\dis {a'\over a_0'}\right)^{\chi^2/ 8}\\
 \where\qquad r_\pm={3\pm\sqrt{9+12k^2}\over k^2}\; , \qquad r_-<\o\xi<r_+,
\end{array}
\ee
and $a_0'$ is an integration constant. For $P'$ of the form (\ref{st5}), the r.h.s. of Eq. (\ref{e3}) vanishes and yields $\dot\xi={c_\xi\over a^{\prime 3}b}$, where $c_\xi\neq 0$ is a constant, so that
\be
\label{mod}
\o \xi={c_\xi\over \dot a'}\, {e^{-\xi}\over a^{\prime 3}}.
\ee
 As follows from Eq. (\ref{sol}) and shown  in Fig. \ref{attraction},
\begin{figure}[h!]
\begin{center}
\vspace{.3cm}
\includegraphics[height=4.5cm]{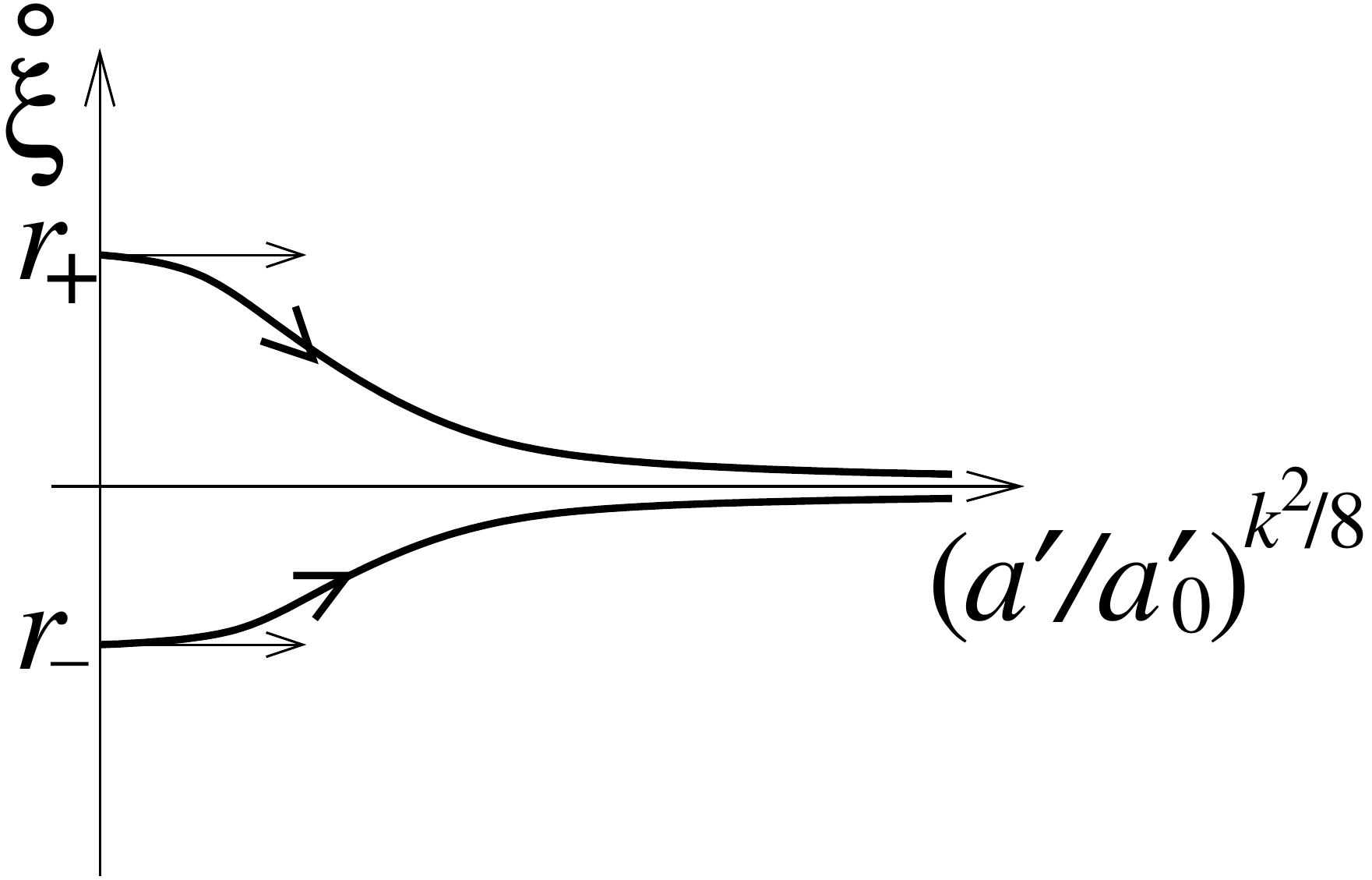}
\caption{\footnotesize \em   $\o\xi={c_\xi\over \dot a'}{e^{-\xi}\over a^{\prime 3}}$ as a function of $(a'/a_0')^{k^2/8}$. For $c_\xi>0$, the expanding solutions correspond to the positive branch, while for $c_\xi<0$ they follow the negative one. In both cases, the scale factor tends to infinity for late times and $\o\xi$ goes to zero. (The indicated tangents can be either horizontal or vertical.)  }
\vspace{-.4cm}
\label{attraction}
\end{center}
\end{figure}
this quantity versus $\dis\left({a'/a_0'}\right)^{\chi^2/ 8}$ has two branches. The expanding solutions have $\dot a>0$ and are described by the positive (negative) branch when $c_\xi>0$ $(c_\xi<0)$. In both cases, we observe that when $t$ increases, the scale factor is monotonic and diverges to $+\infty$, while $(\o\xi,\o\phi{}'_\bot)\to( 0,0)$.

- When $\chi=0$, the solution of Eq. (\ref{eq1}) is
\be
\label{az}
\o\xi={2\over s\left(\dis {a'\over a_0'}\right)^{3/2}-1}\qquad \where \qquad s=\sign\, {\o\xi}\, .
\ee
Using Eq. (\ref{mod}), one obtains
\be
\dot a' \, e^{\xi}={c_\xi\over 2\, a_0^{\prime 3}}\, {s\left(\dis {a'\over a_0'}\right)^{3/2}-1\over \left(\dis {a'\over a_0'}\right)^3}\, .
\ee
The expanding solutions have $\dot a'>0$ and thus $s=\sign\, c_\xi$. Drawing  $\dot a'  e^{\xi}$ versus $(a'/a'_0)^{3/2}$, one concludes that $a'(t)$ always diverges when $t$ increases, and Eq. (\ref{az}) implies $(\o \xi,\o\phi{}'_\bot)\to (0,0)$.

\noindent {\small $\bullet$} Finally, for $\o\xi\equiv 0$, $\o\phi_\bot\not\equiv 0$, we only need to consider Eq. (\ref{e4'}) that reduces to
\be
4\, \oo\phi{}'_\bot+\o \phi{}'_\bot \left( 6-{1\over 2}\,\o \phi{}_\bot^{\prime 2}\right)=0,
\ee
and whose solution is
\be
\label{aze}
\o\phi{}_\bot'=\varepsilon \, \sqrt{12}\, {1\over \sqrt{1+\left(\dis {a'\over a'_0}\right)^3}},
\ee
where $\varepsilon=\pm 1$.
Solving Eq. (\ref{e4}) to find $\dis \dot\phi'_\bot={c'_\bot\over a^{\prime 3}b}$ (where $c'_\bot$ is a constant),  Eq. (\ref{aze}) yields
\be
\dot a'={\varepsilon\,  c'_\bot e^{-\xi_0}\over \sqrt{12}\, a_0^{\prime 3}}\, {\sqrt{1+\left(\dis {a'\over a'_0}\right)^3}\over \left(\dis {a'\over a'_0}\right)^3},
\ee
where $\xi_0$ is the constant value of $\xi$. The expanding solutions have $\varepsilon  c'_\bot >0$ and satisfy $a'\to +\infty$ when $t$ increases. In this limit, Eq. (\ref{aze}) implies $(\o\xi,\o\phi{}'_\bot)\to (0,0)$.


\section*{Appendix F}
\renewcommand{\theequation}{F.\arabic{equation}}
\renewcommand{\thesection}{F.}
\setcounter{equation}{0}

In this Appendix, we complete the discussion in Sect. \ref{sectn2to0nonsusydyn}. We consider phase 3 of the thermal effective potential of a radius $R_6$ that is not participating in the breaking of supersymmetry. We argue that with initial data satisfying  $R_6 > R_0$ and $R_4$, the resulting dynamics implies the evolution is attracted to phase 2 \ie where $R_0$ and $R_4 > R_6$.
The cases under consideration are given in (\ref{models}). In the regime $R_6 \gg R_0$ and $R_4$,  the untwisted sector of the free energy density is given in Eq. (\ref{Fun}), while the twisted sectors can be found in Eqs (\ref{1})--(\ref{4}).

$\bullet$ The case (I.i) is studied in \cite{BEKP}, when the spectrum satisfies
\be
\mbox{(I$b$.i) in 5 dimensions}~:~ -{1\over 31}<{n^u_V+n^t_V\over n^u_T+n^t_T}<0\, ,
\ee
which is the analogue of the second condition in Eq. (\ref{Iabc}) in five rather than four dimensions. Actually, it is found in Sect. 3.3 of \cite{BEKP} that when one neglects the subdominant terms $e^{3y}$ in Eqs. (\ref{Fun}) and (\ref{1}), an RDS attractor exists in five dimensions for arbitrary $y\equiv y_0$ such that $e^{4y_0}\ll 1$. The direction six is part of the space-time and $R_6(t)$ runs away proportionally to $R_{\rm box}(t)$. However, defining $y:=y_0+\varepsilon_{(y)}$, one finds by taking into account the correction terms $\O(e^{3y_0})$ that $\o \varepsilon_{(y)}>0$ (and of order $e^{4y_0}$). It was concluded that $R_0(t)$ (and similarly $R_4(t)$) are ``catching'' $R_6(t)$. We expect this phenomenon is valid until the Universe is attracted back into phase 2 (it would be interesting to confirm this fact by a numerical analysis as in the pure thermal case in appendix A). Once in phase 2, $\zeta(t)$ freezes, the final space-time dimension is four, and the inequalities (\ref{hyp}) end by being better and better satisfied for all radii, including $R_6$.

This analysis can be generalized when the spectrum satisfies
\be
\mbox{(I$a$.i) in 5 dimensions}~:~ {n^u_V+n^t_V\over n^u_T+n^t_T}\le-{1\over 31}\, ,
\ee
which is the analogue in five dimensions of the first condition (\ref{Iabc}).  One finds that the small terms of order $\O(e^{3y_0})$ are again attracting the Universe towards phase 2. The difference with the case (I$b$.i) above is that along the RDS in phase 3, $z$ is not stabilized but only frozen at any value $z_0$ such that $e^{5z_0}\ll 1$.\footnote{\label{tt}To observe the attraction from phase 3 to phase 2, we take the limit $e^{5z}\ll 1$ in  Eqs  (\ref{Fun}) and (\ref{1}). One finds a dominant term $\O(e^{-y-z})$ plus two subdominant monomials of order $e^{-y+4z}$ and $e^{3y}$, respectively. Neglecting these two terms, an RDS is found. However when these residual forces are taken into account, one finds  $\o \varepsilon_{(y)}= e^{4y_0}{12\over 5}{c_4\over c_6}\left(1+{n^u_V+n^t_V\over n^u_T+n^t_T}\right)-e^{5z_0}{3\over 8}{c_5\over c_6}\left({n^u_V+n^t_V\over n^u_T+n^t_T}+{1\over 31}\right)>0$.} As before, we expect the system will enter phase 2 (as could be checked numerically), where $\zeta$ freezes. However, the final space-time dimension in phase 2 can be four (with $z$ stabilized) or five:
\be
\label{ee}
\mbox{4 dimensions if}~ -{1\over 15}<{n^u_V+n^t_V\over n^u_T+n^t_T}\le-{1\over 31}\; , \quad
\mbox{5 dimensions if}~ ~{n^u_V+n^t_V\over n^u_T+n^t_T}\le-{1\over 15}\, ,
\ee
as a consequence of the dynamical decompactification of the direction 4 in the later case.

$\bullet$ The above analysis can be applied similarly to case (II.i) when $n_V^u<0$ (\ie cases (II$a$.i) and (II$b$.i)). The Universe re-enters phase 2 where $\zeta$ freezes, $z$ is stabilized,  and the final RDS attraction is four-dimensional.

$\bullet$ Let us apply the same techniques to case (I.ii). First, we consider models where
 \be
 \label{az'}
\mbox{(I$b$.ii) in five dimensions}~:~ -{1\over 31}<{n^u_V\over n^u_T}<0\, .
\ee
The difference with the discussion of the case (I$b$.i) is that we have a subdominant term in $-T^{-4}\F$ of order zero in $e^y$, $\kappa(z):= n_T^tf_T(z)+n_V^tf_V(z)$. Moreover, the ``residual force'' induced by $\kappa$ can either be positive or negative when $n_V^t<0$, depending on the value of $z$ (instead of being always positive). This could imply that in some models we may not be attracted back to phase 2. However, reasoning as in Sect. 3.3.3 of \cite{BEKP}, one finds that $\kappa$ induces a back-reaction on $z$ and we expect that the residual force on $y$ will always end by increasing it. In any case,  this fact is always true in the explicit models considered in the present  paper, since the condition (\ref{az'}) implies $n_V^t = 0$ and thus  $\kappa(z)>0$ for all $z$ (see Eqs (\ref{ntu}) and (\ref{ntv})). The Universe is attracted towards phase 2,\footnote{Quantitatively, one finds $\o\varepsilon_{(y)}= e^{y_0} {9\over 16}{n_T^tf_T+n_V^tf_V\over n_T^uf^{(5)}_T+n_V^uf^{(5)}_V}\Big\abs_{z'_c}>0$, where $z'_c$ is the value where $z$ is stabilized when $\kappa$ is neglected in phase 3.} where it enters a four-dimensional RDS.

Again, the analysis can be generalized when we have
\be
\mbox{(I$a$.ii) in five dimensions}~:~ {n^u_V\over n^u_T}\le-{1\over 31}\, .
\ee
As usual, $y<0$ increases\footnote{Reasoning as in footnote \ref{tt}, one finds $\o\varepsilon_{(y)}= -e^{5z_0}{3\over 8}{c_5\over c_6}\left({n^u_V\over n_T^u}+{1\over 31}\right)+  e^{y_0}{183\over 80}{c_5\over c_6}{n^t_T\over n_T^u}>0$.} and we approach phase 2. Once there, the final RDS is either four or five dimensional, as indicated in Eq. (\ref{ee}). Note that our analytic discussion does not cover the situations where $n_V^u+n_V^t<0$, with $n_V^u>0$. However, we expect the attraction from phase 3 to 2 to take place as well.

$\bullet$ The discussion in case (II.ii) for $n_V^u<0$ is as in case (I.ii), with a simpler expression for $\kappa=n_T^t\dis \left(1-{\bar \eta\over 2}\right)$.\footnote{One finds in case (II$b$.ii) $\o\varepsilon_{(y)}= e^{y_0} {9\over 16}{n_T^t\left(1-{\bar\eta\over 2}\right)c_4\over (n_T^uf^{(5)}_T+n_V^uf^{(5)}_V)\abs_{z'_c}}>0$, where $z'_c$ is the value where $z$ is stabilized when $\kappa$ is neglected in phase 3. In case (II$a$.ii), one obtains $\o\varepsilon_{(y)}= -e^{5z_0}{3\over 8}{c_5\over c_6}\left({n^u_V\over n_T^u}+{1\over 31}\right)+  e^{y_0}{171\over 80}{c_4\over c_6}{n^t_T\over n_T^u}\left(1-{\bar\eta\over 2}\right)>0$.}
The conclusions are identical, except that the final RDS in phase 2 is always four dimensional.




\vspace{.3cm}
\begin{thebibliography}{99}

\bibitem{inflation}
  A.~H.~Guth,
  ``The inflationary universe: A possible solution to the horizon and flatness problems,''
  Phys.\ Rev.\  D {\bf 23} (1981) 347;

  A.~D.~Linde,
  ``A new inflationary universe scenario: A possible solution of the horizon,
  flatness, homogeneity, isotropy and primordial monopole problems,''
  Phys.\ Lett.\  B {\bf 108} (1982) 389;

  A.~J.~Albrecht and P.~J.~Steinhardt,
  ``Cosmology for grand unified theories with radiatively induced symmetry
  breaking,''
  Phys.\ Rev.\ Lett.\  {\bf 48} (1982) 1220.

\bibitem{stringinflation}
  D.~Baumann and L.~McAllister,
  ``Advances in inflation in string theory,''
  Ann.\ Rev.\ Nucl.\ Part.\ Sci.\  {\bf 59} (2009) 67
  [arXiv:0901.0265 [hep-th]];

  R.~Kallosh,
  ``On inflation in string theory,''
  Lect.\ Notes Phys.\  {\bf 738} (2008) 119
  [arXiv:hep-th/0702059];

  C.~P.~Burgess,
  ``Lectures on cosmic inflation and its potential stringy realizations,''
  PoS {\bf P2GC} (2006) 008
  [Class.\ Quant.\ Grav.\  {\bf 24} (2007\ POSCI,CARGESE2007,003.2007) S795]
  [arXiv:0708.2865 [hep-th]].

\bibitem{cosmomodprob}
  B.~de Carlos, J.~A.~Casas, F.~Quevedo and E.~Roulet,
  ``Model independent properties and cosmological implications of the dilaton and moduli sectors of 4-d strings,''
  Phys.\ Lett.\  B {\bf 318} (1993) 447
  [arXiv:hep-ph/9308325];

  G.~D.~Coughlan, R.~Holman, P.~Ramond and G.~G.~Ross,
  ``Supersymmetry and the entropy crisis,''
  Phys.\ Lett.\  B {\bf 140}, 44 (1984).



\bibitem{AtickWitten}
  J.~Atick and E.~Witten,
  ``The Hagedorn transition and the number of degrees of freedom of string
  theory,''
  Nucl.\ Phys.\  B {\bf 310}, 291 (1988).

\bibitem {AKADK}
  I.~Antoniadis and C.~Kounnas,
  ``Superstring phase transition at high temperature,''
  Phys.\ Lett.\  B {\bf 261} (1991) 369;

  I.~Antoniadis, J.~P.~Derendinger and C.~Kounnas,
  ``Non-perturbative temperature instabilities in $\N = 4$ strings,''
  Nucl.\ Phys.\  B {\bf 551} (1999) 41
  [arXiv:hep-th/9902032];


  C.~Kounnas,
 ``Universal thermal instabilities and the high-temperature phase of the  $\N =
  4$ superstrings,''
  arXiv:hep-th/9902072.

\bibitem{Hagedorn}
  R.~Hagedorn,
 ``Statistical thermodynamics of strong interactions at high-energies,''
  Nuovo Cim.\ Suppl.\  {\bf 3}, 147 (1965);

  S.~Fubini and G.~Veneziano,
  ``Level structure of dual-resonance models,''
  Nuovo Cim.\  A {\bf 64}, 811 (1969);

  K.~Bardakci and S.~Mandelstam,
  ``Analytic solution of the linear-trajectory bootstrap,''
  Phys.\ Rev.\  {\bf 184}, 1640 (1969);

  K.~Huang and S.~Weinberg,
  ``Ultimate temperature and the early universe,''
  Phys.\ Rev.\ Lett.\  {\bf 25}, 895 (1970);

  B.~Sathiapalan,
  ``Vortices on the string world sheet and constraints on toral
  compactification,''
  Phys.\ Rev.\  D {\bf 35}, 3277 (1987);

  Y.~I.~Kogan,
  ``Vortices on the world sheet and string's critical dynamics,''
  JETP Lett.\  {\bf 45}, 709 (1987)
  [Pisma Zh.\ Eksp.\ Teor.\ Fiz.\  {\bf 45}, 556 (1987)];

  M.~Axenides, S.~D.~Ellis and C.~Kounnas,
 ``Universal behavior of $D$-dimensional superstring models,''
  Phys.\ Rev.\  D {\bf 37}, 2964 (1988);

  D.~Kutasov and N.~Seiberg,
  ``Number of degrees of freedom, density of states and tachyons in string
  theory and CFT,''
  Nucl.\ Phys.\  B {\bf 358} (1991) 600;

  D.~Israel and V.~Niarchos,
  ``Tree-level stability without spacetime fermions: Novel examples in string
  theory,''
  JHEP {\bf 0707}, 065 (2007)
  [arXiv:0705.2140 [hep-th]].

\bibitem{hagedorncosmology}
  R.~H.~Brandenberger,
  ``Alternatives to cosmological inflation,''
  arXiv:0902.4731 [hep-th];

  R.~H.~Brandenberger, A.~Nayeri, S.~P.~Patil and C.~Vafa,
  ``String gas cosmology and structure formation,''
  Int.\ J.\ Mod.\ Phys.\  A {\bf 22} (2007) 3621
  [arXiv:hep-th/0608121];

  A.~Nayeri, R.~H.~Brandenberger and C.~Vafa,
  ``Producing a scale-invariant spectrum of perturbations in a Hagedorn phase of string cosmology,''
  Phys.\ Rev.\ Lett.\  {\bf 97} (2006) 021302
  [arXiv:hep-th/0511140];

 N.~Kaloper, L.~Kofman, A.~D.~Linde and V.~Mukhanov,
  ``On the new string theory inspired mechanism of generation of  cosmological perturbations,''
  JCAP {\bf 0610}, 006 (2006)
  [arXiv:hep-th/0608200].

\bibitem{Kaloper:2007pw}
  N.~Kaloper and S.~Watson,
  ``Geometric precipices in string cosmology,''
  Phys.\ Rev.\  D {\bf 77}, 066002 (2008)
  [arXiv:0712.1820 [hep-th]].

 \bibitem{GravFluxes}
 C.~Angelantonj, C.~Kounnas, H.~Partouche and N.~Toumbas,
  ``Resolution of Hagedorn singularity in superstrings with gravito-magnetic
  fluxes,''
  Nucl.\ Phys.\  B {\bf 809} (2009) 291
  [arXiv:0808.1357 [hep-th]].

\bibitem{MassSusy}
  C.~Kounnas,
  ``Massive boson-fermion degeneracy and the early structure of the Universe,''
  Fortsch.\ Phys.\  {\bf 56} (2008) 1143
  [arXiv:0808.1340 [hep-th]];

  I.~Florakis and C.~Kounnas,
  ``Orbifold symmetry reductions of massive boson-fermion degeneracy,''
  Nucl.\ Phys.\  B {\bf 820} (2009) 237
  [arXiv:0901.3055 [hep-th]];
  
  I.~Florakis,
  ``String Models with Massive boson-fermion Degeneracy,''
  Fortsch.\ Phys.\  {\bf 58} (2010) 883
  [arXiv:1001.2589 [hep-th]].

  I.~Florakis, C.~Kounnas and N.~Toumbas,
  ``Marginal Deformations of Vacua with Massive boson-fermion Degeneracy Symmetry,''
  Nucl.\ Phys.\  B {\bf 834} (2010) 273
  [arXiv:1002.2427 [hep-th]].

\bibitem{hybrid}
  I.~Florakis, C.~Kounnas, H.~Partouche and N.~Toumbas,
  ``Non-singular string cosmology in a 2d Hybrid model,''
  Nucl.\ Phys.\  B {\bf 844} (2011) 89
  [arXiv:1008.5129 [hep-th]].

\bibitem{stringgascosmo}
 N.~Matsuo,
  ``Superstring thermodynamics and its application to cosmology,''
  Z.\ Phys.\  C {\bf 36} (1987) 289;

 J.~Kripfganz and H.~Perlt,
  ``Cosmological impact of winding strings,''
  Class.\ Quant.\ Grav.\  {\bf 5} (1988) 453;

  R.~H.~Brandenberger and C.~Vafa,
  ``Superstrings in the early universe,''
  Nucl.\ Phys.\  B {\bf 316} (1989) 391;

  M.~J.~Bowick and S.~B.~Giddings,
  ``High temperature strings,''
  Nucl.\ Phys.\  B {\bf 325}, 631 (1989);

 A.~Tseytlin and C.~Vafa,
  ``Elements of string cosmology,''
  Nucl.\ Phys.\  B {\bf 372} (1992) 443
  [arXiv:hep-th/9109048];

  R.~H.~Brandenberger,
  ``String gas cosmology,''
  arXiv:0808.0746 [hep-th];

  T.~Battefeld and S.~Watson,
  ``String gas cosmology,''
  Rev.\ Mod.\ Phys.\  {\bf 78}, 435 (2006)
  [arXiv:hep-th/0510022];

  B.~A.~Bassett, M.~Borunda, M.~Serone and S.~Tsujikawa,
  ``Aspects of string-gas cosmology at finite temperature,''
  Phys.\ Rev.\  D {\bf 67} (2003) 123506
  [arXiv:hep-th/0301180];

  M.~Borunda and L.~Boubekeur,
  ``The effect of $\alpha'$ corrections in string gas cosmology,''
  JCAP {\bf 0610} (2006) 002
  [arXiv:hep-th/0604085].



\bibitem{KP2}
  C.~Kounnas and H.~Partouche,
  ``Inflationary de Sitter solutions from superstrings,''
  Nucl.\ Phys.\  B {\bf 795} (2008) 334
  [arXiv:0706.0728 [hep-th]];

  C.~Kounnas and H.~Partouche,
  ``Instanton transition in thermal and moduli deformed de Sitter cosmology,''
  Nucl.\ Phys.\  B {\bf 793} (2008) 131
  [arXiv:0705.3206 [hep-th]].

\bibitem{cosmo1}
  T.~Catelin-Jullien, C.~Kounnas, H.~Partouche and N.~Toumbas,
  ``Thermal/quantum effects and induced superstring cosmologies,''
  Nucl.\ Phys.\  B {\bf 797} (2008) 137
  [arXiv:0710.3895 [hep-th]].

\bibitem{cosmo2}
  T.~Catelin-Jullien, C.~Kounnas, H.~Partouche and N.~Toumbas,
  ``Induced superstring cosmologies and moduli stabilization,''
    Nucl.\ Phys.\  B {\bf 820} (2009) 290
  [arXiv:0901.0259 [hep-th]];

  T.~Catelin-Jullien, C.~Kounnas, H.~Partouche and N.~Toumbas,
  ``Thermal and quantum superstring cosmologies,''
  Fortsch.\ Phys.\  {\bf 56} (2008) 792
  [arXiv:0803.2674 [hep-th]].

\bibitem{Bourliot:2009cx}
  F.~Bourliot, J.~Estes, C.~Kounnas and H.~Partouche,
  ``Thermal and quantum induced early superstring cosmology,''
  arXiv:0910.2814 [hep-th].

\bibitem{SS}
  J.~Scherk and J.~H.~Schwarz,
  ``Spontaneous breaking of supersymmetry through dimensional reduction,''
  Phys.\ Lett.\  B {\bf 82} (1979) 60.

\bibitem{Rohm}
  R.~Rohm,
  ``Spontaneous supersymmetry breaking in supersymmetric string theories,''
  Nucl.\ Phys.\  B {\bf 237} (1984) 553.


\bibitem{KouPor}
  C.~Kounnas and M.~Porrati,
  ``Spontaneous supersymmetry breaking in string theory,''
  Nucl.\ Phys.\  B {\bf 310} (1988) 355;

S.~Ferrara, C.~Kounnas, M.~Porrati and F.~Zwirner,
  ``Superstrings with spontaneously broken supersymmetry and their effective theories,''
  Nucl.\ Phys.\  B {\bf 318} (1989) 75.

\bibitem{RostKounnas}
  C.~Kounnas and B.~Rostand,
 ``Coordinate dependent compactifications and discrete symmetries,''
  Nucl.\ Phys.\  B {\bf 341} (1990) 641.

\bibitem{GeoFluxes}
J.~P.~Derendinger, C.~Kounnas, P.~M.~Petropoulos and F.~Zwirner,
  ``Superpotentials in IIA compactifications with general fluxes,''
  Nucl.\ Phys.\  B {\bf 715} (2005) 211
  [arXiv:hep-th/0411276];

  J.~P.~Derendinger, C.~Kounnas, P.~M.~Petropoulos and F.~Zwirner,
  ``Fluxes and gaugings: $\N = 1$ effective superpotentials,''
  Fortsch.\ Phys.\  {\bf 53} (2005) 926
  [arXiv:hep-th/0503229];


  G.~Villadoro and F.~Zwirner,
  ``$\N = 1$ effective potential from dual type-IIA D6/O6 orientifolds with
  general fluxes,''
  JHEP {\bf 0506}, 047 (2005)
  [arXiv:hep-th/0503169];

  G.~Villadoro and F.~Zwirner,
  ``$D$ terms from D-branes, gauge invariance and moduli stabilization in  flux
  compactifications,''
  JHEP {\bf 0603}, 087 (2006)
  [arXiv:hep-th/0602120];

L.~Andrianopoli, M.~A.~Lledo and M.~Trigiante,
  ``The Scherk-Schwarz mechanism as a flux compactification with internal
  torsion,''
  JHEP {\bf 0505} (2005) 051
  [arXiv:hep-th/0502083];

G.~Dall'Agata and N.~Prezas,
  ``Scherk-Schwarz reduction of M-theory on $G_2$-manifolds with fluxes,''
  JHEP {\bf 0510} (2005) 103
  [arXiv:hep-th/0509052];

  J.~P.~Derendinger, P.~M.~Petropoulos and N.~Prezas,
  ``Axionic symmetry gaugings in $\N = 4$ supergravities and their
  higher-dimensional origin,''
  Nucl.\ Phys.\  B {\bf 785}, 115 (2007)
  [arXiv:0705.0008 [hep-th]];

  G.~Curio, A.~Klemm, D.~Lust and S.~Theisen,
  ``On the vacuum structure of type II string compactifications on  Calabi-Yau
  spaces with $H$-fluxes,''
  Nucl.\ Phys.\  B {\bf 609} (2001) 3
  [arXiv:hep-th/0012213].

 \bibitem{OpenFluxes}
  C.~Angelantonj, S.~Ferrara and M.~Trigiante,
  ``New $D = 4$ gauged supergravities from $\N = 4$ orientifolds with fluxes,''
  JHEP {\bf 0310}, 015 (2003)
  [arXiv:hep-th/0306185];

  C.~Angelantonj, R.~D'Auria, S.~Ferrara and M.~Trigiante,
  ``$K3 \times T^2/\Z_2$ orientifolds with fluxes, open string moduli and  critical
 points,''
  Phys.\ Lett.\  B {\bf 583}, 331 (2004)
  [arXiv:hep-th/0312019];

  C.~Angelantonj, S.~Ferrara and M.~Trigiante,
  ``Unusual gauged supergravities from type IIA and type IIB orientifolds,''
  Phys.\ Lett.\  B {\bf 582}, 263 (2004)
  [arXiv:hep-th/0310136];

  C.~Angelantonj, M.~Cardella and N.~Irges,
  ``An alternative for moduli stabilisation,''
  Phys.\ Lett.\  B {\bf 641} (2006) 474
  [arXiv:hep-th/0608022].

\bibitem{Noscale}
  E.~Cremmer, S.~Ferrara, C.~Kounnas and D.~V.~Nanopoulos,
  ``Naturally vanishing cosmological constant in $\N=1$ supergravity,''
  Phys.\ Lett.\  B {\bf 133} (1983) 61;

  J.~R.~Ellis, C.~Kounnas and D.~V.~Nanopoulos,
  ``No scale supersymmetric GUTs,''
  Nucl.\ Phys.\  B {\bf 247} (1984) 373;

  J.~R.~Ellis, C.~Kounnas and D.~V.~Nanopoulos,
  ``Phenomenological $SU(1,1)$ supergravity,''
  Nucl.\ Phys.\  B {\bf 241} (1984) 406;

  J.~R.~Ellis, A.~B.~Lahanas, D.~V.~Nanopoulos and K.~Tamvakis,
 ``No-scale supersymmetric standard model,''
  Phys.\ Lett.\  B {\bf 134}, 429 (1984).

\bibitem{StringyNoscale}
  E.~Witten,
  ``Dimensional reduction of superstring models,''
  Phys.\ Lett.\  B {\bf 155} (1985) 151;

  S.~Ferrara, C.~Kounnas and M.~Porrati,
  ``General dimensional reduction of ten-dimensional supergravity and
  superstring,''
  Phys.\ Lett.\  B {\bf 181} (1986) 263;

 M.~Cvetic, J.~Louis and B.~A.~Ovrut,
  ``A string calculation of the K\"ahler potentials for moduli of $\Z_N$
  orbifolds,''
  Phys.\ Lett.\  B {\bf 206} (1988) 227;

  L.~J.~Dixon, V.~Kaplunovsky and J.~Louis,
  ``On effective field theories describing $(2,2)$ vacua of the heterotic
  string,''
  Nucl.\ Phys.\  B {\bf 329} (1990) 27;

 M.~Cvetic, J.~Molera and B.~A.~Ovrut,
  ``K\"ahler potentials for matter scalars and moduli of $\Z_N$ orbifolds,''
  Phys.\ Rev.\  D {\bf 40} (1989) 1140.

\bibitem{BEKP}
  F.~Bourliot, J.~Estes, C.~Kounnas and H.~Partouche,
``Cosmological phases of the string thermal effective potential,''
  Nucl.\ Phys.\  B {\bf 830} (2010) 330
  [arXiv:0908.1881 [hep-th]].


\bibitem{Partouche:2010cq}
  H.~Partouche,
  ``Attractions to radiation-like eras in superstring cosmologies,''
  Fortsch.\ Phys.\  {\bf 58} (2010) 797
  [arXiv:1003.0840 [hep-th]].



\bibitem{Watson:2004aq}
  S.~Watson,
  ``Moduli stabilization with the string Higgs effect,''
  Phys.\ Rev.\  D {\bf 70}, 066005 (2004)
  [arXiv:hep-th/0404177].

\bibitem{Patil:2004zp}
  S.~P.~Patil and R.~Brandenberger,
  ``Radion stabilization by stringy effects in general relativity and dilaton gravity,''
  Phys.\ Rev.\  D {\bf 71}, 103522 (2005)
  [arXiv:hep-th/0401037].

\bibitem{Greene:2007sa}
  B.~Greene, S.~Judes, J.~Levin, S.~Watson and A.~Weltman,
  ``Cosmological moduli dynamics,''
  JHEP {\bf 0707} (2007) 060
  [arXiv:hep-th/0702220].

\bibitem{Bourliot:2009na}
  F.~Bourliot, C.~Kounnas and H.~Partouche,
  ``Attraction to a radiation-like era in early superstring cosmologies,''
  Nucl.\ Phys.\  B {\bf 816} (2009) 227
  [arXiv:0902.1892 [hep-th]].


\bibitem{NoscaleTSR}
  S.~Ferrara, C.~Kounnas and F.~Zwirner,
  ``Mass formulae and natural hierarchy in string effective supergravities,''
  Nucl.\ Phys.\  B {\bf 429} (1994) 589
  [Erratum-ibid.\  B {\bf 433} (1995) 255]
  [arXiv:hep-th/9405188];

  C.~Kounnas, F.~Zwirner and I.~Pavel,
  ``Towards a dynamical determination of parameters in the minimal
  supersymmetric standard model,''
  Phys.\ Lett.\  B {\bf 335} (1994) 403
  [arXiv:hep-ph/9406256];

  C.~Kounnas, I.~Pavel, G.~Ridolfi and F.~Zwirner,
  ``Possible dynamical determination of $m_t$, $m_b$ and $m_\tau$,''
  Phys.\ Lett.\  B {\bf 354} (1995) 322
  [arXiv:hep-ph/9502318].

\bibitem{Assel:2010wj}
  B.~Assel, K.~Christodoulides, A.~E.~Faraggi, C.~Kounnas and J.~Rizos,
  ``Classification of heterotic Pati-Salam models,''
  arXiv:1007.2268 [hep-th];

  B.~Assel, K.~Christodoulides, A.~E.~Faraggi, C.~Kounnas and J.~Rizos,
  ``Exophobic quasi-realistic heterotic string vacua,''
  Phys.\ Lett.\  B {\bf 683} (2010) 306
  [arXiv:0910.3697 [hep-th]].

\bibitem{Antoniadis:1986ke}
  I.~Antoniadis and C.~Kounnas,
 ``The dilaton classical solution and the supersymmetry breaking evolution in
  an expanding universe,''
  Nucl.\ Phys.\  B {\bf 284} (1987) 729.

\bibitem{Estes:2011iw}
  J.~Estes, L.~Liu, H.~Partouche,
  ``Massless D-strings and moduli stabilization in type I cosmology,''
  [arXiv:1102.5001 [hep-th]].

\bibitem{KTT}
C.~Kounnas, N.~Toumbas and J.~Troost,
  ``A wave-function for stringy universes,''
  JHEP {\bf 0708} (2007) 018
  [arXiv:0704.1996 [hep-th]].


\bibitem{intradcor}
  L.~J.~Dixon, V.~Kaplunovsky and J.~Louis,
  ``Moduli dependence of string loop corrections to gauge coupling constants,''
  Nucl.\ Phys.\  B {\bf 355} (1991) 649;

  E.~Kiritsis, C.~Kounnas, P.~M.~Petropoulos and J.~Rizos,
  ``On the heterotic effective action at one-loop, gauge couplings and the gravitational sector,''
  arXiv:hep-th/9605011;

  E.~Kiritsis, C.~Kounnas, P.~M.~Petropoulos and J.~Rizos,
  Nucl.\ Phys.\  B {\bf 540} (1999) 87
  [arXiv:hep-th/9807067].

\bibitem{het/II}
  A.~Strominger,
  ``Massless black holes and conifolds in string theory,''
  Nucl.\ Phys.\  B {\bf 451} (1995) 96
  [arXiv:hep-th/9504090];

  S.~Kachru and C.~Vafa,
  ``Exact results for $\N=2$ compactifications of heterotic strings,''
  Nucl.\ Phys.\  B {\bf 450} (1995) 69
  [arXiv:hep-th/9505105];

  S.~Kachru, A.~Klemm, W.~Lerche, P.~Mayr and C.~Vafa,
  ``Nonperturbative results on the point particle limit of $\N=2$ heterotic string
  compactifications,''
  Nucl.\ Phys.\  B {\bf 459} (1996) 537
  [arXiv:hep-th/9508155];

  I.~Antoniadis and H.~Partouche,
  ``Exact monodromy group of $\N=2$ heterotic superstring,''
  Nucl.\ Phys.\  B {\bf 460} (1996) 470
  [arXiv:hep-th/9509009];

  H.~Partouche,
  ``Non perturbative check of $\N = 2$, $D = 4$ heterotic/type II duality,''
  Nucl.\ Phys.\ Proc.\ Suppl.\  {\bf 55B} (1997) 210
  [arXiv:hep-th/9610119].

\bibitem{KP}
  P.~Kaste and H.~Partouche,
  ``On the equivalence of $\N = 1$ brane worlds and geometric singularities  with
  flux,''
  JHEP {\bf 0411} (2004) 033
  [arXiv:hep-th/0409303].

\bibitem{Gregori:1997hi}
  A.~Gregori, E.~Kiritsis, C.~Kounnas, N.~A.~Obers, P.~M.~Petropoulos and B.~Pioline,
  ``$R^2$ corrections and non-perturbative dualities of $\N = 4$ string ground
  states,''
  Nucl.\ Phys.\  B {\bf 510} (1998) 423
  [arXiv:hep-th/9708062].



\end{thebibliography}
\end{document}